\documentstyle[epsfig]{mn2e}

\begin{document}

\title[WiggleZ Survey: growth of structure]{The WiggleZ Dark Energy
  Survey: the growth rate of cosmic structure since redshift $z =
  0.9$}

\author[Blake et al.]{\parbox[t]{\textwidth}{Chris Blake$^1$, Sarah
    Brough$^2$, Matthew Colless$^2$, Carlos Contreras$^1$, Warrick
    Couch$^1$, Scott Croom$^3$, Tamara Davis$^{4,5}$, Michael
    J.\ Drinkwater$^4$, Karl Forster$^6$, David Gilbank$^7$, Mike
    Gladders$^8$, Karl Glazebrook$^1$, Ben Jelliffe$^3$, Russell
    J.\ Jurek$^9$, I-hui Li$^1$, Barry Madore$^{10}$, D.\ Christopher
    Martin$^6$, Kevin Pimbblet$^{11}$, Gregory B.\ Poole$^1$, Michael
    Pracy$^{1,12}$, Rob Sharp$^{2,12}$, Emily Wisnioski$^1$, David
    Woods$^{13}$, Ted K.\ Wyder$^6$ and H.K.C. Yee$^{14}$} \\ \\ $^1$
  Centre for Astrophysics \& Supercomputing, Swinburne University of
  Technology, P.O. Box 218, Hawthorn, VIC 3122, Australia \\ $^2$
  Australian Astronomical Observatory, P.O. Box 296, Epping, NSW 1710,
  Australia \\ $^3$ Sydney Institute for Astronomy, School of Physics,
  University of Sydney, NSW 2006, Australia \\ $^4$ School of
  Mathematics and Physics, University of Queensland, Brisbane, QLD
  4072, Australia \\ $^5$ Dark Cosmology Centre, Niels Bohr Institute,
  University of Copenhagen, Juliane Maries Vej 30, DK-2100 Copenhagen
  \O, Denmark \\ $^6$ California Institute of Technology, MC 278-17,
  1200 East California Boulevard, Pasadena, CA 91125, United States
  \\ $^7$ Astrophysics and Gravitation Group, Department of Physics
  and Astronomy, University of Waterloo, Waterloo, ON N2L 3G1, Canada
  \\ $^8$ Department of Astronomy and Astrophysics, University of
  Chicago, 5640 South Ellis Avenue, Chicago, IL 60637, United States
  \\ $^9$ Australia Telescope National Facility, CSIRO, Epping, NSW
  1710, Australia \\ $^{10}$ Observatories of the Carnegie Institute
  of Washington, 813 Santa Barbara St., Pasadena, CA 91101, United
  States \\ $^{11}$ School of Physics, Monash University, Clayton, VIC
  3800, Australia \\ $^{12}$ Research School of Astronomy \&
  Astrophysics, Australian National University, Weston Creek, ACT
  2600, Australia \\ $^{13}$ Department of Physics \& Astronomy,
  University of British Columbia, 6224 Agricultural Road, Vancouver,
  BC V6T 1Z1, Canada \\ $^{14}$ Department of Astronomy and
  Astrophysics, University of Toronto, 50 St.\ George Street, Toronto,
  ON M5S 3H4, Canada}

\maketitle

\begin{abstract}
We present precise measurements of the growth rate of cosmic structure
for the redshift range $0.1 < z < 0.9$, using redshift-space
distortions in the galaxy power spectrum of the WiggleZ Dark Energy
Survey.  Our results, which have a precision of around $10\%$ in four
independent redshift bins, are well-fit by a flat $\Lambda$CDM
cosmological model with matter density parameter $\Omega_{\rm m} =
0.27$.  Our analysis hence indicates that this model provides a
self-consistent description of the growth of cosmic structure through
large-scale perturbations and the homogeneous cosmic expansion mapped
by supernovae and baryon acoustic oscillations.  We achieve robust
results by systematically comparing our data with several different
models of the quasi-linear growth of structure including empirical
models, fitting formulae calibrated to N-body simulations, and
perturbation theory techniques.  We extract the first measurements of
the power spectrum of the velocity divergence field,
$P_{\theta\theta}(k)$, as a function of redshift (under the assumption
that $P_{g\theta}(k) = -\sqrt{P_{gg}(k) P_{\theta\theta}(k)}$ where
$g$ is the galaxy overdensity field), and demonstrate that the WiggleZ
galaxy-mass cross-correlation is consistent with a deterministic
(rather than stochastic) scale-independent bias model for WiggleZ
galaxies for scales $k < 0.3 \, h$ Mpc$^{-1}$.  Measurements of the
cosmic growth rate from the WiggleZ Survey and other current and
future observations offer a powerful test of the physical nature of
dark energy that is complementary to distance-redshift measures such
as supernovae and baryon acoustic oscillations.
\end{abstract}
\begin{keywords}
surveys, large-scale structure of Universe, cosmological parameters
\end{keywords}

\section{Introduction}
\renewcommand{\thefootnote}{\fnsymbol{footnote}}
\setcounter{footnote}{1}
\footnotetext{E-mail: cblake@astro.swin.edu.au}

Recent cosmological observations have revealed significant gaps in
our understanding of the physics of the Universe.  A set of
measurements including the anisotropies of the Cosmic Microwave
Background radiation, the shape of the clustering power spectrum of
galaxies, the brightness of distant supernovae and the projected
scales of baryon acoustic oscillations have indicated the presence of
a ``dark energy'' component which is propelling the cosmic expansion
into a phase of acceleration (for recent results see Komatsu et
al.\ 2009, Reid et al.\ 2009, Percival et al.\ 2010, Guy et
al.\ 2010).

The physical nature of dark energy is not yet understood.  Several
explanations have been put forward including the presence of
smoothly-distributed energy such as a cosmological constant or a
quintessence scalar field, a large-scale modification to Einstein's
theory of General Relativity, or the effects of spatially-varying
curvature in an inhomogeneous Universe.  Further observational data is
required to distinguish clearly between the subtly-varying predictions
of these very different physical models (e.g., Linder 2005, Wang
2008, Wiltshire 2009).

One of the most important observational datasets for addressing this
issue is the large-scale structure of the galaxy distribution.  The
clustering within this distribution arises through a process of
gravitational instability which acts to amplify primordial matter
fluctuations.  The growth rate of this structure with time is a key
discriminant between cosmological models (e.g., Linder \& Jenkins
2003, Linder \& Cahn 2007, Nesseris \& Perivolaropoulos 2008).  Two
different physical dark energy scenarios with the same background
cosmic expansion generally produce different growth rates of
perturbations, hence growth measurements are able to discriminate
between models that are degenerate under geometric tests (Davis et
al.\ 2007, Rubin et al.\ 2009).

The growth of cosmic structure is driven by the motion of matter, for
which galaxies act as ``tracer particles''.  These flows imprint a
clear observational signature in galaxy surveys, known as
redshift-space distortions, because the galaxy redshift is generated
by not only the background cosmic expansion but also the peculiar
velocity tracing the bulk flow of matter (Kaiser 1987, Hamilton 1998).
As a consequence the 2-point statistics of the galaxy distribution are
anisotropic on large scales, where the amplitude of the anisotropy is
related to the velocity of the bulk flow and hence to the growth rate
of structure.

Many previous galaxy surveys have measured this anisotropy employing
either the galaxy correlation function or power spectrum.  In the
relatively local Universe, exquisite studies at redshift $z \approx
0.1$ have been undertaken using data from the 2-degree Field Galaxy
Redshift Survey (2dFGRS; Peacock et al.\ 2001, Hawkins et al.\ 2003,
Percival et al.\ 2004) and the Sloan Digital Sky Survey (SDSS; Tegmark
et al.\ 2004).  The SDSS Luminous Red Galaxy (LRG) sample enabled
these measurements to be extended to somewhat higher redshifts $z
\approx 0.35$ (Tegmark et al.\ 2006, Cabre \& Gaztanaga 2009, Okumura
et al.\ 2008).  Noisier results have been achieved at greater
look-back times up to $z \approx 1$ by surveys mapping significantly
smaller cosmic volumes such as the 2dF-SDSS LRG and Quasar survey
(2SLAQ; da Angela et al.\ 2008) and the VIMOS-VLT Deep Survey (VVDS;
Guzzo et al.\ 2008).

The current observational challenge is to map the
intermediate-redshift $0.3 < z < 1$ Universe with surveys of
comparable statistical power to 2dFGRS and SDSS, so that accurate
growth rate measurements can be extracted across the full
(hypothesized) redshift range for which dark energy dominates the
cosmic dynamics.  In order to achieve this goal we have constructed
the WiggleZ Dark Energy Survey (Drinkwater et al.\ 2010), a new
large-scale spectroscopic galaxy redshift survey, using the
multi-object AAOmega fibre spectrograph at the $3.9$m Australian
Astronomical Telescope.  The survey, which began in August 2006,
targets UV-selected star-forming emission-line galaxies in several
different regions around the sky and at redshifts $z < 1$.  By
covering a total solid angle of $1000$ deg$^2$ the WiggleZ Survey maps
two orders of magnitude more cosmic volume in the $z > 0.5$ Universe
than previous galaxy redshift surveys.  This paper presents the
current measurements of the growth rate of structure using the galaxy
power spectrum of the survey.  The dataset will also permit many other
tests of the cosmological model via baryon acoustic oscillations, the
Alcock-Pacyznski effect, higher-order clustering statistics and
topological descriptors of the density field.

The improving statistical accuracy with which redshift-space
distortions may be measured by observational datasets requires that
the theoretical modelling of the data also advances.  Recent reviews
of this topic have been provided by Percival \& White (2009) and Song
\& Percival (2009).  In the linear clustering regime, in the
large-scale limit, the theory is well-understood (Kaiser 1987,
Hamilton 1998).  However, both simulations and observations have
demonstrated that linear theory is a poor approximation across a wide
range of quasi-linear scales encoding a great deal of clustering
information (e.g.\ Jennings et al.\ 2011, Okumura et al.\ 2011).  The
blind application of linear-theory modelling to current surveys would
therefore result in a significant systematic error in the extraction
of the growth rate and a potential mis-diagnosis of the physical
nature of dark energy.

Various methodologies have been employed for extending the modelling
of redshift-space distortions to quasi-linear and non-linear scales.
The classical approach -- the so-called ``streaming model''
(e.g.\ Hatton \& Cole 1998) -- modulates the linear theory clustering
with an empirical damping function characterized by a variable
parameter, which is marginalized over when extracting the growth rate.
This model has provided an acceptable statistical fit to many previous
datasets, but the lack of a strong physical motivation for the
empirical function could lead to systematic errors when the model is
confronted by high-precision data.

In this paper we consider two alternatives.  Firstly, quasi-linear
redshift-space distortions can be modelled by various
physically-motivated perturbation theory schemes (for recent reviews
see Nishimichi et al. 2009; Carlson, White \& Padmanabhan 2010).
Given that the accuracy of some current perturbation techniques breaks
down at a particular quasi-linear scale, leading to potentially large
discrepancies at smaller scales, the range of validity of these models
must be carefully considered.  The second approach is to use numerical
N-body simulations to produce fitting formulae for the density and
velocity power spectra (Smith et al.\ 2003, Jennings et al.\ 2011),
which enables models to be established across a wider range of scales.
The main concern of this approach is that these fitting formulae may
only be valid for the subset of cosmologies and galaxy formation
models in which they were derived (an important point given the
unknown nature of dark energy).

A further significant issue in the modelling of redshift-space
distortions in the galaxy distribution is the ``galaxy bias'' relation
by which galaxies trace the matter overdensities that drive the
velocities (e.g., Cole \& Kaiser 1989).  The typical assumption of a
local, linear, deterministic bias, for which there is a good physical
motivation on large scales (Scherrer \& Weinberg 1998), may break down
in the case of precise measurements of the clustering pattern on
quasi-linear scales (Swanson et al.\ 2008), also potentially leading
to systematic errors in growth rate fits.  In this study we consider
the introduction of stochasticity to the bias relation by varying the
galaxy-mass cross-correlation (Dekel \& Lahav 1999).  We note that
further studies of the WiggleZ dataset involving the bispectrum,
3-point correlation function, galaxy halo occupation distribution and
comparison with numerical simulations will yield further constraints
on the form of galaxy bias.

The aim of this paper is to use the existing range of redshift-space
distortion models and galaxy power spectra from the WiggleZ survey to
derive measurements of the growth rate across the redshift range $z <
1$ that are robust against modelling systematics.  We assume
throughout a cosmological model consistent with the analysis of the
latest measurements of the CMB by the Wilkinson Microwave Anisotropy
Probe (Komatsu et al.\ 2009): a flat Universe described by General
Relativity with matter density $\Omega_{\rm m} = 0.27$, cosmological
constant $\Omega_\Lambda = 0.73$, baryon fraction $\Omega_{\rm
  b}/\Omega_{\rm m} = 0.166$, Hubble parameter $h = 0.72$, primordial
scalar index of fluctuations $n_{\rm s} = 0.96$ and total fluctuation
amplitude $\sigma_8 = 0.8$.  In addition to providing a good
description of the temperature and polarization fluctuations in the
CMB, this model yields a good fit to distance measurements from
supernovae and baryon acoustic oscillations (Guy et al.\ 2010,
Percival et al.\ 2010).  In this paper we test if the same model also
predicts the observed growth rate of structure.  Future studies will
explore simultaneous fits to these datasets using different dark
energy models.

The layout of this paper is as follows: in Section \ref{secmeas} we
present the measurements of the various observational statistics
quantifying the anisotropic power spectrum.  Section \ref{secmod}
summarizes the current theory of redshift-space distortions in Fourier
space and introduces in more detail the models we will fit to the
data.  In Section \ref{secparfit} we carry out the parameter fitting
focussing on the growth rate and the galaxy-mass cross-correlation.
Section \ref{secpkmom} presents an analysis of the moments of the
power spectrum and extraction of the power spectrum of the velocity
divergence field, and Section \ref{secconc} lists our conclusions.

\section{Measurements}
\label{secmeas}

\begin{figure*}
\center
\epsfig{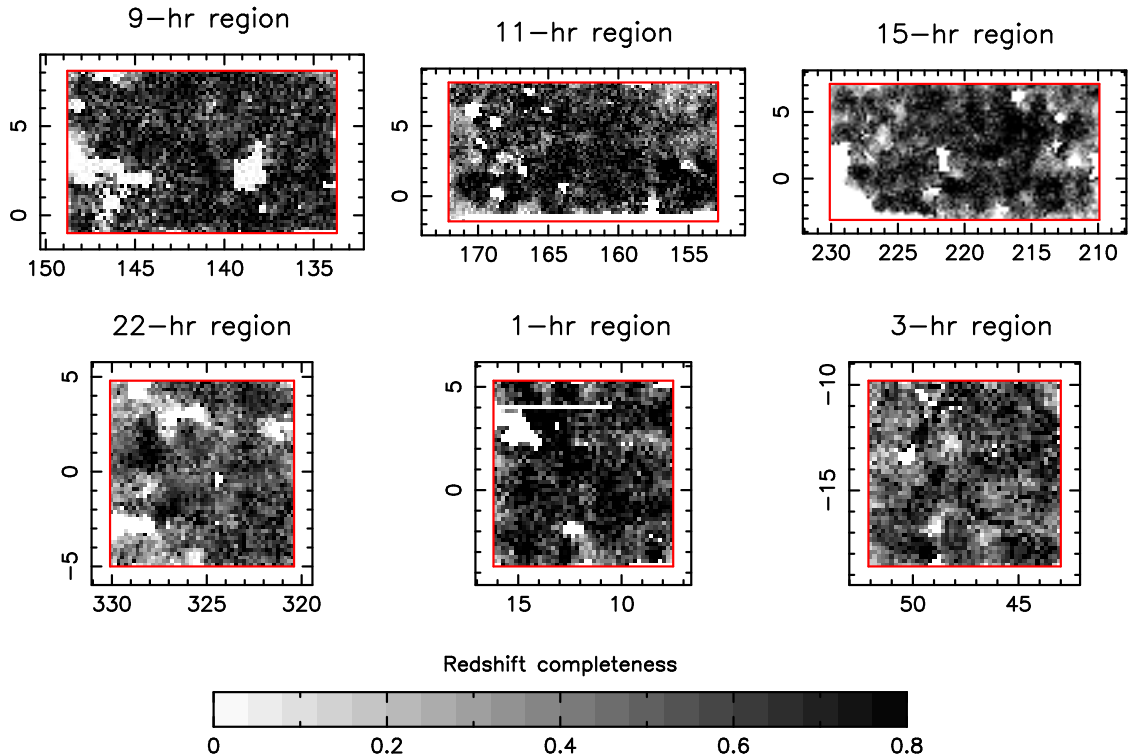}
\caption{Greyscale map illustrating the relative redshift completeness
  of each of the six WiggleZ survey regions analyzed in this paper.
  This Figure is generated by taking the ratio of the galaxy densities
  in the redshift and parent catalogues in small cells.  The $x$-axis
  and $y$-axis of each panel represent right ascension and
  declination, respectively.}
\label{figradec}
\end{figure*}

\subsection{Sample}

The WiggleZ Dark Energy Survey at the Australian Astronomical
Telescope (Drinkwater et al. 2010) is a large-scale galaxy redshift
survey of bright emission-line galaxies mapping a cosmic volume of
order 1 Gpc$^3$ over redshift $z < 1$.  The survey, which began in
August 2006 and is scheduled to finish in January 2011, will obtain of
order $200{,}000$ redshifts for UV-selected galaxies covering of order
1000 deg$^2$ of equatorial sky, using the AAOmega spectrograph (Sharp
et al.\ 2006).  The survey design is driven by the scientific goal of
measuring baryon acoustic oscillations in the galaxy power spectrum at
a significantly higher redshift than existing surveys.  The target
galaxy population is selected from UV imaging by the Galaxy Evolution
Explorer (GALEX) satellite, matched with optical data from the Sloan
Digital Sky Survey (SDSS) and Red Cluster Sequence survey (RCS2;
Gilbank et al.\ 2011) to provide an accurate position for fibre
spectroscopy.  Full details about the survey design, execution and
modelling are provided by Blake et al.\ (2009), Drinkwater et
al.\ (2010) and Blake et al.\ (2010).

In this paper we analyze a subset of the WiggleZ sample assembled up
to the end of the 10A semester (May 2010).  We include data from six
survey regions -- the 9-hr, 11-hr, 15-hr, 22-hr, 1-hr and 3-hr regions
-- in the redshift range $0.1 < z < 0.9$, which together constitute a
total sample of $N = 152{,}117$ galaxies.  Figure \ref{figradec}
displays the distribution in right ascension and declination of the
analyzed sample for the six survey regions, where the greyscale level
corresponds to the relative redshift completeness.  We divided the
sample into four redshift slices of width $\Delta z = 0.2$ in order to
map the evolution of the growth rate with redshift.  The effective
redshifts at which the clustering pattern was measured in each of
these slices (evaluated using equation 44 of Blake et al.\ 2010) were
$z_{\rm eff} = (0.22, 0.41, 0.60, 0.78)$.  The numbers of galaxies
analyzed in each redshift slice were $N = (19608, 39495, 60227,
32787)$.

\subsection{Power spectrum estimation}

We estimated the two-dimensional galaxy power spectrum $P_g(k,\mu)$ in
four redshift slices for each of the six WiggleZ survey regions using
the direct Fourier methods introduced by Feldman, Kaiser \& Peacock
(1994; FKP).  Our methodology is fully described in Section 3.1 of
Blake et al.\ (2010); we give a brief summary here.  Firstly we mapped
the angle-redshift survey cone into a cuboid of co-moving co-ordinates
using a fiducial flat $\Lambda$CDM cosmological model with matter
density $\Omega_{\rm m} = 0.27$.  We gridded the catalogue in cells
using nearest grid point assignment ensuring that the Nyquist
frequencies in each direction were much higher than the Fourier
wavenumbers of interest (we corrected the power spectrum measurement
for the small bias introduced by this gridding using the method of
Jing 2005).  We then applied a Fast Fourier Transform to the grid.
The window function of each region was determined using the methods
described by Blake et al.\ (2010) that model effects due to the survey
boundaries, incompleteness in the parent UV and optical catalogues,
incompleteness in the spectroscopic follow-up, systematic variations
in the spectroscopic redshift completeness across the AAOmega
spectrograph, and variations of the galaxy redshift distribution with
angular position.  The Fast Fourier Transform of the window function
was then used to construct the final power spectrum estimator.  The
measurement was corrected for the small effect of redshift blunders
using Monte Carlo survey simulations as described in Section 3.2 of
Blake et al.\ (2010).

Since each WiggleZ survey region subtends a relatively small angle on
the sky, of order 10 degrees, the flat-sky approximation is valid.  We
orient the $x$-axis of our Fourier cuboid parallel to the
line-of-sight at the angular centre of each region, and then represent
each Fourier mode by wavevectors parallel and perpendicular to the
line-of-sight: $k_\parallel = |k_x|$ and $k_\perp = \sqrt{k_y^2 +
  k_z^2}$.  We can also then determine the values of the total
wavenumber of each mode $k = \sqrt{k_\parallel^2 + k_\perp^2}$ and the
cosine of its angle to the line-of-sight, $\mu = k_\parallel/k$.  We
used two binning schemes for averaging the Fourier modes $\vec{k}$: in
bins of $k_\perp$ and $k_\parallel$ (of width $0.02 \, h$ Mpc$^{-1}$)
and in bins of $k$ and $\mu$ (of width $0.02 \, h$ Mpc$^{-1}$ and
$0.1$, respectively).  We determined the covariance matrix of the
power spectrum measurement in these binning schemes by implementing
the sums in Fourier space described by FKP (see Blake et al.\ 2010
equations 20-22).  The angular size of each WiggleZ region implies
that the effect of wide-angle distortions (Raccanelli, Samushia \&
Percival 2010) is not significant.

We note that the FKP covariance matrix of the power spectrum only
includes the contribution from the survey window function and neglects
any covariance due to non-linear growth of structure or redshift-space
effects.  The full covariance matrix may be studied with the aid of a
large ensemble of N-body simulations (Rimes \& Hamilton 2005,
Takahashi et al.\ 2011), which we are preparing in conjunction with
the final WiggleZ survey sample.  The impact of using the full
non-linear covariance matrix on growth-of-structure measurements has
not yet been studied, although Takahashi et al.\ (2011) demonstrated
that the effect on the accuracy of extraction of the baryon acoustic
oscillations is very small.

The power spectrum model must be convolved with the window function to
be compared to the data.  For reasons of computing speed we re-cast
the convolution as a matrix multiplication
\begin{equation}
P_{\rm convolved}(i) = \sum_j M_{ij} \, P_{\rm model}(j) \; ,
\label{eqconv}
\end{equation}
where $i$ and $j$ label a single bin in the two-dimensional set
$(k_\perp,k_\parallel)$ or $(k,\mu)$.  We determined the convolution
matrix $M_{ij}$ by evaluating the full Fourier convolution for a
complete set of unit vectors.  For example, to evaluate the $j^{\rm
  th}$ row of matrix elements, corresponding to a bin $(k_{{\rm
    min},j} < k < k_{{\rm max},j}, \mu_{{\rm min},j} < \mu < \mu_{{\rm
    max},j})$, we defined the three-dimensional model in Fourier space
for the unit vector
\begin{eqnarray}
P_{\rm model}(\vec{k}) &=& 1 \hspace{5mm} (k_{{\rm min},j} < k < k_{{\rm max},j} ; \mu_{{\rm min},j} < \mu < \mu_{{\rm max},j}) , \nonumber \\
&=& 0 \hspace{5mm} {\rm otherwise} \; ,
\end{eqnarray}
applied the full convolution transform (equation 16 in Blake et
al.\ 2010), and binned the resulting power spectrum amplitudes in the
same $(k,\mu)$ bins.  The vector of results defines the $j^{\rm th}$
row of the matrix $M$ in Equation \ref{eqconv}.  In summary, for each
of the 24 sub-regions we obtain a data vector $P_g^s$ [spanning
  $(k_\perp,k_\parallel)$ or $(k,\mu)$], a covariance matrix and a
convolution matrix.

\begin{figure*}
\center
\epsfig{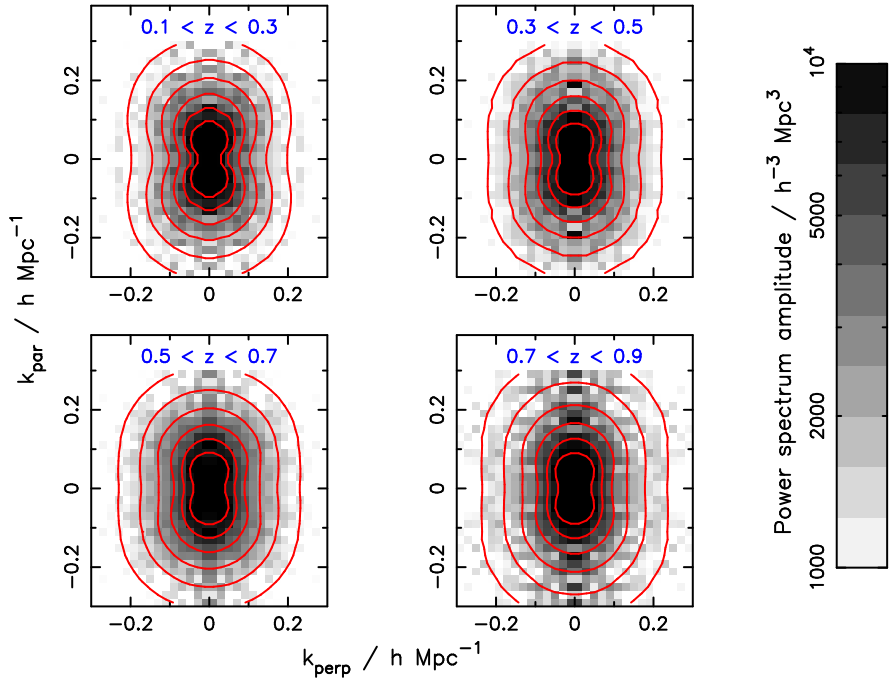}
\caption{The galaxy power spectrum amplitude as a function of
  wavevectors $(k_\perp,k_\parallel)$ perpendicular and parallel to
  the line-of-sight, determined by stacking observations in different
  WiggleZ survey regions in four redshift slices.  The contours
  correspond to the best-fitting non-linear empirical Lorentzian
  redshift-space distortion model.  We note that because of the
  differing degrees of convolution in each region due to the window
  function, a ``de-convolution'' method was used to produce this plot.
  Before stacking, the data points were corrected by the ratio of the
  unconvolved and convolved two-dimensional power spectra
  corresponding to the best-fitting model, for the purposes of this
  visualization.  Only the top-right quadrant of data for each
  redshift is independent; the other three quadrants are mirrors of
  this first quadrant.  The $k_\perp = 0$ axis is noisiest because it
  contains the lowest number of Fourier modes available for power
  spectrum determination.}
\label{figpkcomp}
\end{figure*}

\begin{figure*}
\center
\epsfig{file=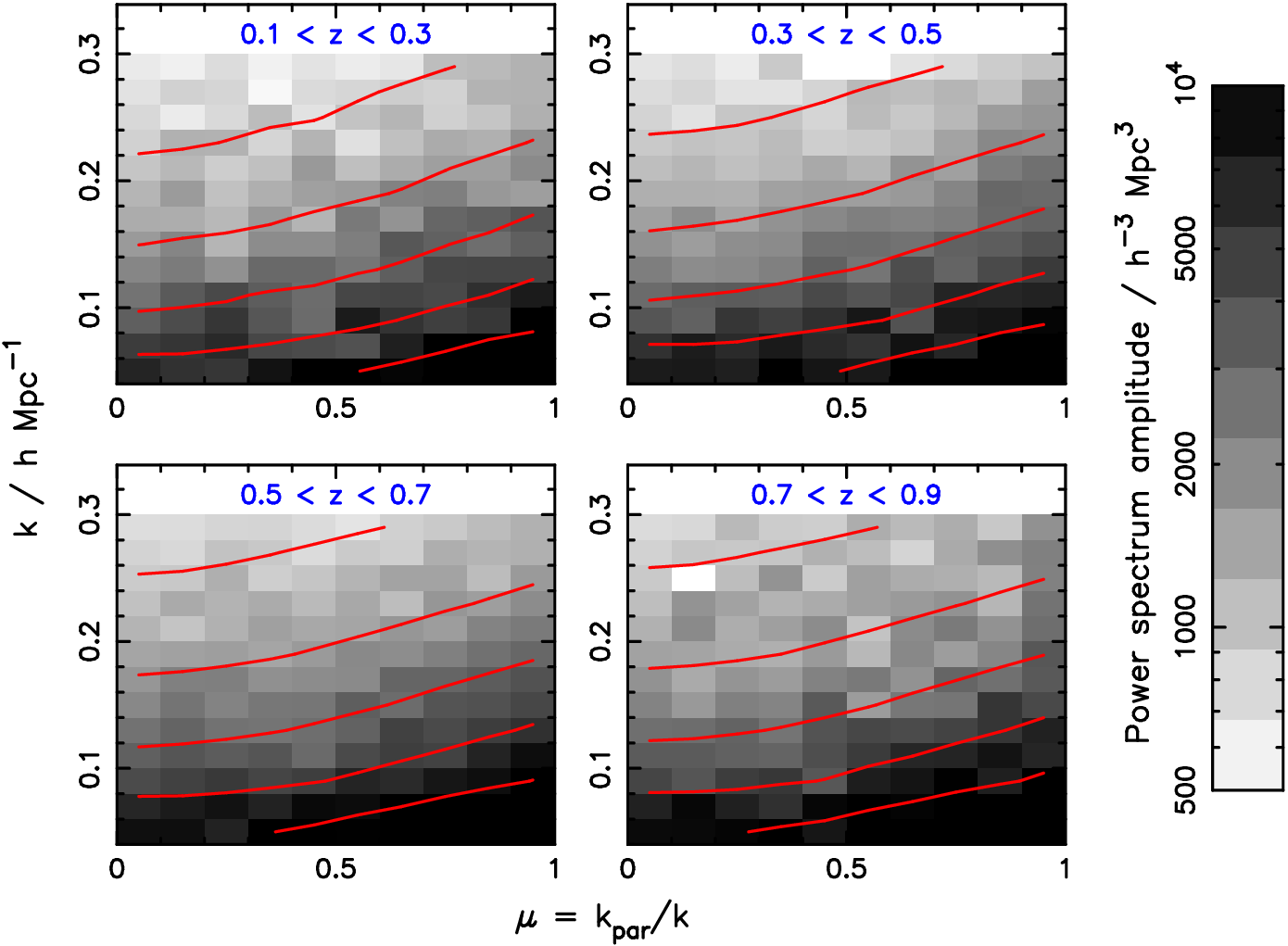,width=9cm,angle=-90}
\caption{The galaxy power spectrum as a function of amplitude and
  angle of Fourier wavevector $(k,\mu)$, determined by stacking
  observations in different WiggleZ survey regions in four redshift
  slices.  The contours correspond to the best-fitting non-linear
  empirical Lorentzian redshift-space distortion model.  A similar
  stacking method was used to that employed in the generation of
  Figure \ref{figpkcomp}.  In the absence of redshift-space
  distortions, the model contours would be horizontal lines.}
\label{figpkmu}
\end{figure*}

Figures \ref{figpkcomp} and \ref{figpkmu} respectively display
two-dimensional power spectra $P(k_\perp,k_\parallel)$ and $P(k,\mu)$
for each of the four redshift slices, obtained by stacking
measurements across the six survey regions.  For comparison we also
plot in each case contours corresponding to the best-fitting
non-linear empirical Lorentzian redshift-space distortion model
described below.  In Figure \ref{figpkcomp} the non-circular nature of
the measurements and models in Fourier space encode the imprint of
redshift-space distortions.  The overall ``squeezing'' of the contours
in the $k_\perp$ direction reflects the large-scale bulk flows.  The
apparent ``pinching'' of the models near the $k_\parallel = 0$ axis is
due to the damping caused by the pairwise velocity dispersion
discussed below in Section \ref{secempmodel}, the amplitude of which
is seen to increase with decreasing redshift (the pinching results
from the relative variation with $\mu$ of the numerator and
denominator of Equation \ref{eqpkemp2}).  Figure \ref{figpkmu}, which
bins the clustering amplitude with the cosine of the angle to the
line-of-sight $\mu$, illustrates how the coherent velocity flows boost
the power of radial ($\mu=1$) modes relative to tangential ($\mu=0$)
modes for a given scale $k$.

\section{Modelling the redshift-space galaxy power spectrum}
\label{secmod}

\begin{figure*}
\center
\epsfig{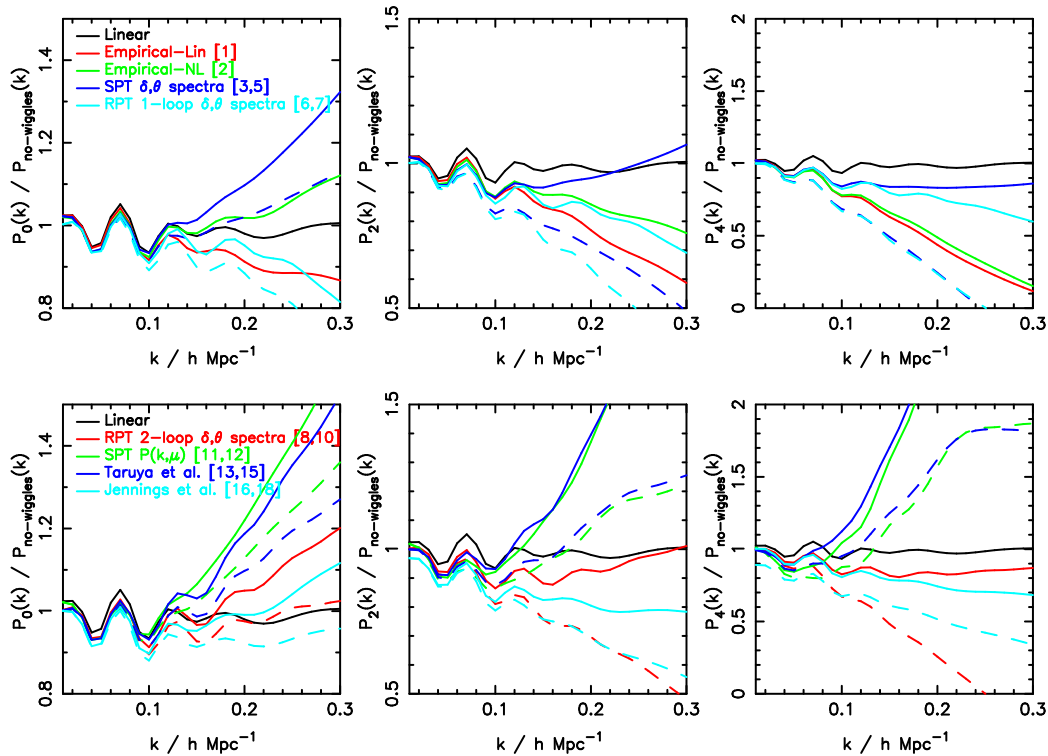}
\caption{The multipole power spectra $P_\ell(k)$ for $\ell=0,2,4$ for
  the different models listed in Table \ref{tabmodlist}.  The models
  are evaluated at redshift $z=0.6$ assuming a linear bias $b=1$, a
  growth rate $f=0.7$ and (where applicable) a damping term $\sigma_v
  = 300 \, h$ km s$^{-1}$.  The models are labelled by their row
  number in Table \ref{tabmodlist}.  The solid and dashed lines are
  models that respectively include and exclude the damping term.  All
  models are divided by a smooth ``no-wiggles'' reference power
  spectrum from the fitting formulae of Eisenstein \& Hu (1998), which
  has the same shape as the linear power spectrum but without the
  imprint of baryon acoustic oscillations.  The models agree well in
  the large-scale limit, but significant differences develop between
  the models at smaller scales.}
\label{figpklmod}
\end{figure*}

In this Section we describe a range of 18 models of the redshift-space
galaxy power spectrum in the quasi-linear regime that we will try
fitting to our measurements.  These models are listed in Table
\ref{tabmodlist}.  We assume that the shape of the underlying linear
matter power spectrum is accurately determined by observations of the
Cosmic Microwave Background radiation, and hence we fix the background
cosmological parameters.  In this case each redshift-space power
spectrum model contains at least two parameters to be fitted: the
growth rate $f$ and a linear bias $b$.  In several cases discussed
below we introduce a third parameter, a variable damping coefficient
$\sigma_v$.  The multipole power spectra of these models at redshift
$z=0.6$ are compared in Figure \ref{figpklmod} assuming a linear bias
$b=1$, a growth rate $f=0.7$ and (where applicable) a damping term
$\sigma_v = 300 \, h$ km s$^{-1}$.  For the purposes of illustration,
all models in Figure \ref{figpklmod} are divided by a smooth
``no-wiggles'' reference power spectrum from the fitting formulae of
Eisenstein \& Hu (1998), which has the same shape as the linear power
spectrum but without the imprint of baryon acoustic oscillations.

\begin{table*}
\begin{center}
\caption{Description of the quasi-linear redshift-space power spectrum
  models fitted to the WiggleZ survey measurements to determine the
  growth rate $f$.  The ``Damping'' for each model can be ``Variable''
  (empirically fit to the data), ``Linear'' (determined using
  Equation \ref{eqsig} as motivated by Scoccimarro 2004) or ``None''
  (not included in the model).  In each model we also fit for a linear
  bias parameter $b$.}
\begin{tabular}{ccccc}
\hline
& Model & Damping & Fitted parameters & Reference \\
\hline
1. & Empirical Lorentzian with linear $P_{\delta\delta}(k)$ & Variable & $f$, $b$, $\sigma_v$ & e.g.\ Hatton \& Cole (1998) \\
2. & Empirical Lorentzian with non-linear $P_{\delta\delta}(k)$ & Variable & $f$, $b$, $\sigma_v$ & \\
3. & $P_{\delta\delta}$, $P_{\delta\theta}$, $P_{\theta\theta}$ from 1-loop SPT & None & $f$, $b$ & e.g.\ Vishniac (1983), Juszkiewicz et al.\ (1984) \\
4. & $P_{\delta\delta}$, $P_{\delta\theta}$, $P_{\theta\theta}$ from 1-loop SPT & Variable & $f$, $b$, $\sigma_v$ & \\
5. & $P_{\delta\delta}$, $P_{\delta\theta}$, $P_{\theta\theta}$ from 1-loop SPT & Linear & $f$, $b$ & \\
6. & $P_{\delta\delta}$, $P_{\delta\theta}$, $P_{\theta\theta}$ from 1-loop RPT & None & $f$, $b$ & Crocce \& Scoccimarro (2006) \\
7. & $P_{\delta\delta}$, $P_{\delta\theta}$, $P_{\theta\theta}$ from 1-loop RPT & Linear & $f$, $b$ & \\
8. & $P_{\delta\delta}$, $P_{\delta\theta}$, $P_{\theta\theta}$ from 2-loop RPT & None & $f$, $b$ & \\
9. & $P_{\delta\delta}$, $P_{\delta\theta}$, $P_{\theta\theta}$ from 2-loop RPT & Variable & $f$, $b$ & \\
10. & $P_{\delta\delta}$, $P_{\delta\theta}$, $P_{\theta\theta}$ from 2-loop RPT & Linear & $f$, $b$ & \\
11. & $P(k,\mu)$ from 1-loop SPT & None & $f$, $b$ & Matsubara (2008) \\
12. & $P(k,\mu)$ from 1-loop SPT & Linear & $f$, $b$ & \\
13. & $P(k,\mu)$ with additional corrections & None & $f$, $b$ & Taruya et al.\ (2010) \\
14. & $P(k,\mu)$ with additional corrections & Variable & $f$, $b$, $\sigma_v$ & \\
15. & $P(k,\mu)$ with additional corrections & Linear & $f$, $b$ & \\
16. & Fitting formulae from N-body simulations & None & $f$, $b$ & Smith et al.\ (2003), Jennings et al.\ (2011) \\
17. & Fitting formulae from N-body simulations & Variable & $f$, $b$, $\sigma_v$ & \\
18. & Fitting formulae from N-body simulations & Linear & $f$, $b$ & \\
\hline
\label{tabmodlist}
\end{tabular}
\end{center}
\end{table*}

\subsection{Density and velocity power spectra}

The galaxy overdensity field, $\delta_g$, is modified in
redshift-space by peculiar velocities.  In Fourier space the
redshift-space overdensity field is given by
\begin{equation}
\delta_g^s(k,\mu) = \delta_g(k) - \mu^2 \theta(k) \; ,
\label{eqdelta}
\end{equation}
where $\theta(k)$ is the Fourier transform of the divergence of the
peculiar velocity field $\vec{u}$ in units of the co-moving Hubble
velocity (i.e.\ $\vec{u} = \vec{v}/[H(a) a]$), $\theta = \vec{\nabla}
. \vec{u}$, and $\mu$ is the cosine of the angle of the Fourier mode
to the line-of-sight.  Equation \ref{eqdelta} assumes that the galaxy
separation is small compared with the distance to the galaxies,
$\delta_g$ and $\theta$ are small, the velocity field $\vec{u}$ is
irrotational, and the continuity equation holds.  In this case the
linear redshift-space power spectrum of a population of galaxies may
be written
\begin{equation}
P_g^s(k,\mu) = P_{gg}(k) - 2 \mu^2 P_{g\theta}(k) + \mu^4 P_{\theta\theta}(k) \; ,
\label{eqpklin}
\end{equation}
where $P_{gg}(k) \equiv \langle |\delta_g(\vec{k})|^2 \rangle$,
$P_{g\theta}(k) \equiv \langle \delta_g(\vec{k}) \theta^*(\vec{k})
\rangle$, $P_{\theta\theta}(k) \equiv \langle |\theta(\vec{k})|^2
\rangle$ are the isotropic galaxy-galaxy, galaxy-$\theta$ and
$\theta$-$\theta$ power spectra for modes $\vec{k}$, respectively.  We
will often refer to the $P_{\theta\theta}(k)$ as the ``velocity power
spectrum'' although it would be better described as the ``power
spectrum of the velocity divergence field''.

Assuming that the velocity field is generated under linear perturbation
theory then
\begin{equation}
\theta(k) = - f \, \delta(k) \; ,
\label{eqflow}
\end{equation}
where $f$ is the growth rate of structure, expressible in terms of the
growth factor $D(a)$ at cosmic scale factor $a$ as $f \equiv
d\ln{D}/d\ln{a}$, and $\delta$ is the matter overdensity.  The growth
factor is defined in terms of the amplitude of a single perturbation
as $\delta(a) = D(a) \, \delta(1)$.  Equation \ref{eqflow}
additionally assumes that the linearized Euler and Poisson equations
hold in a perturbed Friedmann-Robertson-Walker universe.  It
represents a coherent flow of matter in which there is a one-to-one
coupling between the Fourier components of the velocity divergence and
density fields.

Under the assumption of a deterministic, scale-independent, local,
linear bias $b$ then
\begin{equation}
\delta_g = b \, \delta \; ,
\end{equation}
and we may write $P_{gg} = b^2 P_{\delta\delta}$ and $P_{g\theta} = b
\, P_{\delta\theta}$.  If we additionally assume that Equation
\ref{eqflow} applies, then Equation \ref{eqpklin} may be written
\begin{eqnarray}
P_g^s(k,\mu) &=& b^2 \, P_{\delta\delta}(k) \left( 1 + \frac{f \mu^2}{b}
\right)^2 \nonumber \\
&=& b^2 \, P_{\delta\delta}(k) \, (1 + \beta \mu^2)^2 \; .
\label{eqpkkaiser}
\end{eqnarray}
Equation \ref{eqpkkaiser} is known as the large-scale ``Kaiser limit''
of the redshift-space power spectrum model (Kaiser 1987), often
expressed in terms of the parameter $\beta = f/b$.  We assume no
velocity bias between galaxies and matter (Lau, Nagai \& Kravtsov
2010).

Simulations and observations have demonstrated that Equation
\ref{eqpkkaiser} is an unreliable model on all but the largest scales
(smallest values of $k$) due to the non-linear growth of structure.
Deviations from the Kaiser limit are evident for $k > 0.02 \, h$
Mpc$^{-1}$ and are particularly noticeable in the $\theta$ power
spectra (Jennings et al.\ 2011; Okumura et al. 2011).  This failure of
the model is due to the breakdown of the relation between $\theta$ and
$\delta$ (Equation \ref{eqflow}) rather than the underlying structure
of Equation \ref{eqpklin} (Scoccimarro 2004).  Non-linear evolution
implies that a given overdensity $\delta$ produces a range of values
of $\theta$, and this range of velocities acts to smooth the galaxy
overdensity field in redshift-space, or damp the $\theta$ power
spectra.  This non-linear damping must be modelled in order to avoid
introducing a systematic error into our extraction of the growth rate
$f$ from data.  A variety of methods are available for implementing
this non-linear correction, which we discuss below.

\subsection{The empirical non-linear velocity model}
\label{secempmodel}

The standard ``streaming model'' for describing the non-linear
component of redshift-space distortions (e.g.\ Hatton \& Cole 1998)
introduces an empirical damping function $F$ to be multiplied into
Equation \ref{eqpklin}, representing convolution with uncorrelated
galaxy motions on small scales:
\begin{equation}
P_g^s(k,\mu) = \left[ P_{gg}(k) - 2 \mu^2 P_{g\theta}(k) + \mu^4
  P_{\theta \theta}(k) \right] F(k,\mu) \; .
\label{eqpkstream}
\end{equation}
The two models most commonly considered in the literature are the
Lorentzian $F = [1 + (k \sigma_v \mu)^2]^{-1}$ and the Gaussian $F =
\exp{[-(k \sigma_v \mu)^2]}$, representing exponential and Gaussian
convolutions in configuration space, and each parameterized by a
single variable $\sigma_v$.  The Lorentzian model produces better fits
to data (e.g.\ Hawkins et al.\ 2003, Cabre \& Gaztanaga 2009) and we
assume this version of the model in our study.

Equation \ref{eqpkstream} is typically applied assuming that $P_{gg}$,
$P_{g\theta}$ and $P_{\theta\theta}$ are predicted by linear theory
assuming Equation \ref{eqflow}, hence for the Lorentzian model we
obtain
\begin{equation}
P_g^s(k,\mu) = b^2 \, P_{\delta\delta,{\rm lin}}(k) \frac{(1 + f
  \mu^2/b)^2}{1 + (k \sigma_v \mu)^2} \; ,
\label{eqpkemp1}
\end{equation}
where we generated the linear power spectrum $P_{\delta\delta,{\rm
    lin}}(k)$ using the CAMB software package (Lewis, Challinor \&
Lasenby 2000).  This is {\it Model 1} in Table \ref{tabmodlist}.  We
also considered the case where a non-linear density power spectrum
$P_{\delta\delta,{\rm NL}}(k)$, generated by applying the fitting
formula of Smith et al.\ (2003) to the CAMB output, is used in
Equation \ref{eqpkemp1}:
\begin{equation}
P_g^s(k,\mu) = b^2 \, P_{\delta\delta,{\rm NL}}(k) \frac{(1 + f
  \mu^2/b)^2}{1 + (k \sigma_v \mu)^2} \; .
\label{eqpkemp2}
\end{equation}
This is {\it Model 2}.

Although these models are motivated by virialized motions of particles
in collapsed structures, it is important to note that they are
heuristic in nature.  The correction represented by $F(k)$ in fits to
real data is typically of order $20\%$ at $k \sim 0.2 \, h$
Mpc$^{-1}$.  These Fourier modes describe physical scales of tens of
$h^{-1}$ Mpc, far exceeding the size of virialized structures.  In
addition, the form of $F$ and the value of $\sigma_v$ depend strongly
on details such as galaxy type, dark matter halo mass and satellite
fraction.  However, it should be noted that Equation \ref{eqpkemp1}
does a very reasonable job of empirically modelling real datasets at
the precision available in previous redshift surveys (e.g., Hawkins et
al.\ 2003, Cabre \& Gaztanaga 2009).

\subsection{Perturbation theory approaches}

A different approach to modelling clustering beyond linear scales is
to extend Equations \ref{eqpklin} and \ref{eqflow} into the
quasi-linear regime using perturbation theory techniques.  These
approaches have the advantage of a stronger physical motivation
compared to the empirical streaming models, and the disadvantage that
they are potentially applicable for a narrower range of scales,
depending on the type of perturbation expansion.  Standard
perturbation theory at $z = 0$ is only accurate for the range $k < 0.1
\, h$ Mpc$^{-1}$, but other expansion approaches are available with
the precise range of validity dependent on the model in question and
the accuracy required (Nishimichi et al.\ 2009, Carlson et al.\ 2009).
We describe the order of the perturbative expansion by the number of
``loops'' of resummation performed; calculations including up to 2
loops are currently tractable.

Various methodologies have been introduced.  The simplest technique is
to use perturbation theory approaches to model the quasi-linear
behaviour of the functions $P_{\delta\delta}(k)$,
$P_{\delta\theta}(k)$ and $P_{\theta\theta}(k)$ in Equation
\ref{eqpklin}.  These techniques have been recently reviewed by
Nishimichi et al.\ (2009) and Carlson et al.\ (2009) and include
Eulerian standard perturbation theory (SPT; e.g.\ Vishniac 1983,
Juszkiewicz, Sonoda \& Barrow 1984) together with attempts to improve
the convergence behaviour such as Renormalized Perturbation Theory
(RPT; e.g.\ Crocce \& Scoccimarro 2006) which does not expand on the
amplitude of fluctuations.  When generating the perturbation theory
predictions we assumed an input linear power spectrum consistent with
the latest CMB observations: $\Omega_{\rm m} = 0.27$, $\Omega_\Lambda
= 0.73$, $\Omega_{\rm b}/\Omega_{\rm m} = 0.166$, $h = 0.72$, $n_{\rm
  s} = 0.96$ and $\sigma_8 = 0.8$.\footnote{We are very grateful to
  Martin Crocce for providing us with the 1-loop and 2-loop outputs of
  RPT for our cosmological model at the redshifts in question.}

Going beyond the linear assumption may also lead to an alternative
dependence of the redshift-space power spectrum on $\mu$ to that
exhibited by Equation \ref{eqpklin}.  Scoccimarro (2004) proposed the
following model for the redshift-space power spectrum in terms of the
quasi-linear density and velocity power spectra:
\begin{equation}
P_g^s(k,\mu) = \left[ P_{gg}(k) - 2 \mu^2 P_{g\theta}(k) + \mu^4
  P_{\theta\theta}(k) \right] e^{-(k \mu \sigma_v)^2} \; ,
\label{eqdamp}
\end{equation}
where $\sigma_v$ is determined by
\begin{equation}
\sigma_v^2 = \frac{1}{6 \pi^2} \int P_{\theta\theta}(k) \, dk \; .
\label{eqsig}
\end{equation}
The power spectra $P_{gg}(k)$, $P_{g\theta}(k)$ and
$P_{\theta\theta}(k)$ in Equation \ref{eqdamp} may be generated by
perturbation theory or other approaches.  We note that
$P_{g\theta}(k)$ and $P_{\theta\theta}(k)$ are functions of $f$.  The
damping factor in Equation \ref{eqdamp} is analogous to the streaming
model of Equation \ref{eqpkstream} but has a very different physical
motivation: it aims to model the quasi-linear growth of the power
spectra rather than virialized small-scale motions.  Indeed, it would
be possible to add an extra empirical damping factor $F$ to Equation
\ref{eqdamp} to model small-scale motions.

As discussed by Scoccimarro (2004), the model of Equation \ref{eqdamp}
is an approximation in which the Gaussian damping factor attempts to
reproduce the correct non-linear behaviour; Equation \ref{eqdamp}
cannot be strictly derived from theory.  Given this approximation we
consider fitting $\sigma_v$ as a variable parameter in addition to
fixing it using Equation \ref{eqsig}.  {\it Models 3 to 10} in Table
\ref{tabmodlist} are various combinations of SPT and RPT with
different implementations of the damping term.

We note that $\sigma_v$ can also be expressed in velocity units by
multiplying by the Hubble parameter $H_0 = 100 \, h$ km s$^{-1}$
Mpc$^{-1}$.  When calculating the damping term we use the linear
velocity power spectrum as the input to Equation \ref{eqsig}; i.e.\ we
set $P_{\theta\theta}(k) = f^2 P_{\delta\delta,{\rm lin}}(k)$.

The final perturbation theory approaches we consider follow Matsubara
(2008) and Taruya, Nishimichi \& Saito (2010) who present quasi-linear
perturbation theory models including terms up to $\mu^6$, of the form
\begin{equation}
P_g^s(k,\mu) = \sum_{n=0}^3 A_n(k) \, \mu^{2n} \; ,
\end{equation}
where the coefficients $A_n(k)$ are functions of $f$, which we also
try fitting to our data.  The Matsubara (2008) results are a full
angle-dependent treatment of standard perturbation theory {\it (Models
  11 and 12)}, and Taruya et al.\ (2010) present an improved analysis
incorporating additional correction terms {\it (Models 13 to 15)}.
When calculating the Taruya et al.\ model prediction we use
power-spectra $P_{\delta\delta}(k)$, $P_{\delta\theta}(k)$ and
$P_{\theta\theta}(k)$ generated by 2-loop Renormalized Perturbation
Theory.

\subsection{Fitting formulae calibrated by simulations}

Finally, N-body dark matter simulations can be exploited to calibrate
the quasi-linear forms of the functions $P_{\delta\delta}(k)$,
$P_{\delta\theta}(k)$ and $P_{\theta\theta}(k)$.  The advantage of
this technique is that the results will be (potentially) reliable
across a wider range of scales than is accessible with perturbation
theory.  The disadvantage is that simulations are expensive to
generate and it is difficult to span a wide range of input
cosmological models (although that is not a limitation for us given
that we are only considering a single fiducial model).

Smith et al.\ (2003) presented a widely-used prescription for
generating non-linear density power spectra $P_{\delta\delta}$.
Fitting formulae calibrated to N-body simulations for
$P_{\delta\theta}$ and $P_{\theta\theta}$ as a function of redshift,
in terms of $P_{\delta\delta}$, were recently proposed by Jennings et
al.\ (2011).  We inserted these fitting functions into Equation
\ref{eqdamp}, scaling by $f$ and $f^2$ respectively to correct for the
differing notation conventions.  The Jennings et al.\ formulae,
combined with various implementations of the damping term, are {\it
  Models 16 to 18} in Table \ref{tabmodlist}.

\section{Parameter fits}
\label{secparfit}

\subsection{Growth rate}
\label{secgrowthfit}

We fitted the 18 models introduced in Section \ref{secmod} and
summarized in Table \ref{tabmodlist} to the WiggleZ Survey galaxy
power spectra $P_g(k_\perp,k_\parallel)$ measured in Section
\ref{secmeas}.  For each of the four redshift slices we determined the
growth rate $f$ fitting to the six survey regions, marginalizing over
the linear bias $b$ (and the pairwise velocity dispersion $\sigma_v$
where applicable).  We also recorded the minimum value of the $\chi^2$
statistic for each model calculated using the full covariance matrix.
We repeated this procedure varying the range of scales $0 <
\sqrt{k_\perp^2 + k_\parallel^2} < k_{\rm max}$ over which each model
is fitted.  Utilizing a higher value of $k_{\rm max}$ produces an
improved statistical error in the measurement, but potentially causes
a larger systematic error since all models (and particularly some of
the perturbation-theory models) are less reliable at larger values of
$k$ for which the non-linear corrections are more significant.  In the
absence of systematic errors the best-fitting growth rate would be
independent of $k_{\rm max}$.

\begin{figure*}
\center
\epsfig{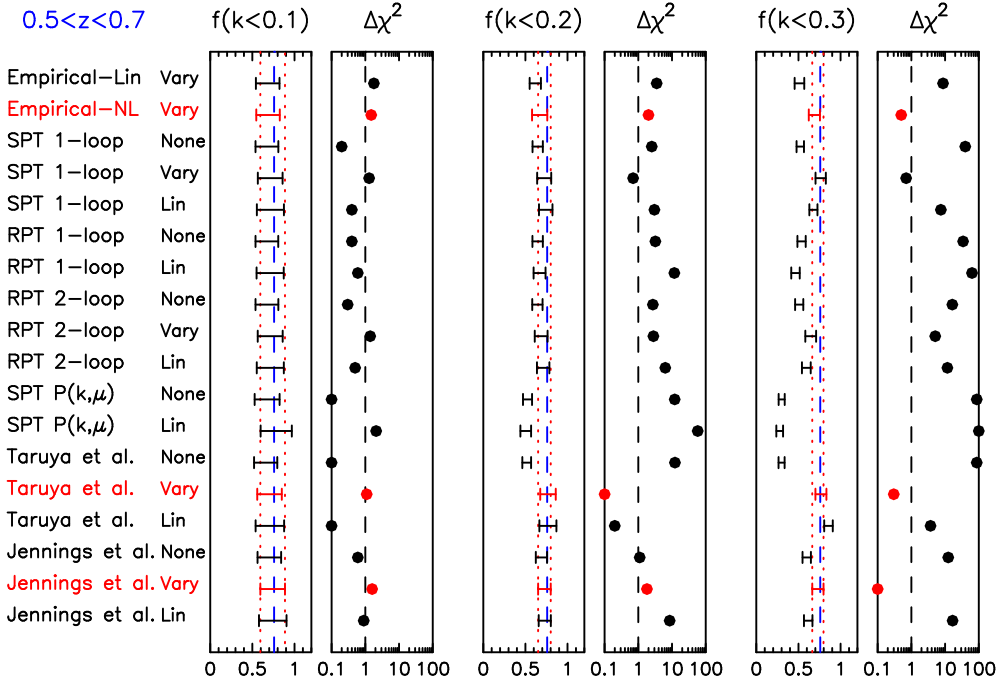}
\caption{Measurements of the growth rate $f$ for the $0.5 < z < 0.7$
  redshift slice for each of the 18 models listed in Table
  \ref{tabmodlist}.  The three panels, each consisting of a pair of
  plots, correspond to different ranges of fitted scales $0 < k <
  k_{\rm max}$ where $k_{\rm max} = 0.1, 0.2$ and $0.3 \, h$
  Mpc$^{-1}$.  For each panel the left-hand plot shows the measurement
  of $f$ and the right-hand plot displays the minimum value of the
  $\chi^2$ statistic relative to the best-fitting model for that
  choice of $k_{\rm max}$.  In the left-hand plot, the vertical dashed
  line indicates the prediction of a flat $\Lambda$CDM cosmological
  model with $\Omega_{\rm m} = 0.27$.  The two vertical dotted lines
  span the $68\%$ confidence region of the growth rate measured for
  the Jennings et al.\ model with a variable damping parameter,
  facilitating an easy comparison of the results for different models.
  In the right-hand plot, points with $\Delta \chi^2 < 0.1$ are
  plotted at the left-hand edge of the panel and $\Delta \chi^2 = 1$
  is indicated by the vertical dashed line.  The three best-performing
  models for $k_{\rm max} = 0.3$ are highlighted by red text.}
\label{figfmod}
\end{figure*}

Figure \ref{figfmod} displays the growth-rate measurements for the
$0.5 < z < 0.7$ redshift slice (which produces the highest statistical
accuracy of the four slices), comparing results for $k_{\rm max} =
0.1$, $0.2$ and $0.3 \, h$ Mpc$^{-1}$.  At least one model can always
be found that provides a good fit to the data for each of the choices
of $k_{\rm max}$, as indicated by the minimum values of $\chi^2 =
(93.8, 436.5, 999.1)$ for $k_{\rm max} = (0.1, 0.2, 0.3)$ with number
of degrees of freedom $(87, 411, 981)$.  The respective probabilties
for obtaining values of $\chi^2$ higher than these are $(0.29, 0.19,
0.34)$, indicating an acceptable goodness-of-fit.  In Figure
\ref{figfmod} we display the minimum values of $\chi^2$ for every
model relative to the best-fitting model for each choice of $k_{\rm
  max}$.

For $k_{\rm max} = 0.1$ all models provide a good fit to the data and
consistent measurements of the growth rate.  This confirms the
convergence of the different modelling approaches at large scales.
For $k_{\rm max} = 0.2$ and $0.3$ some models are significantly
disfavoured by larger values of $\chi^2$, and these models produce
measurements of the growth rate which systematically differ from the
best-performing models.  For $k_{\rm max} = 0.3$, models with a
variable damping parameter produce a fit with a significantly lower
value of $\chi^2$, suggesting that Equation \ref{eqsig} produces an
unreliable prediction of the damping coefficient for these smaller
scales.

\begin{figure*}
\center
\epsfig{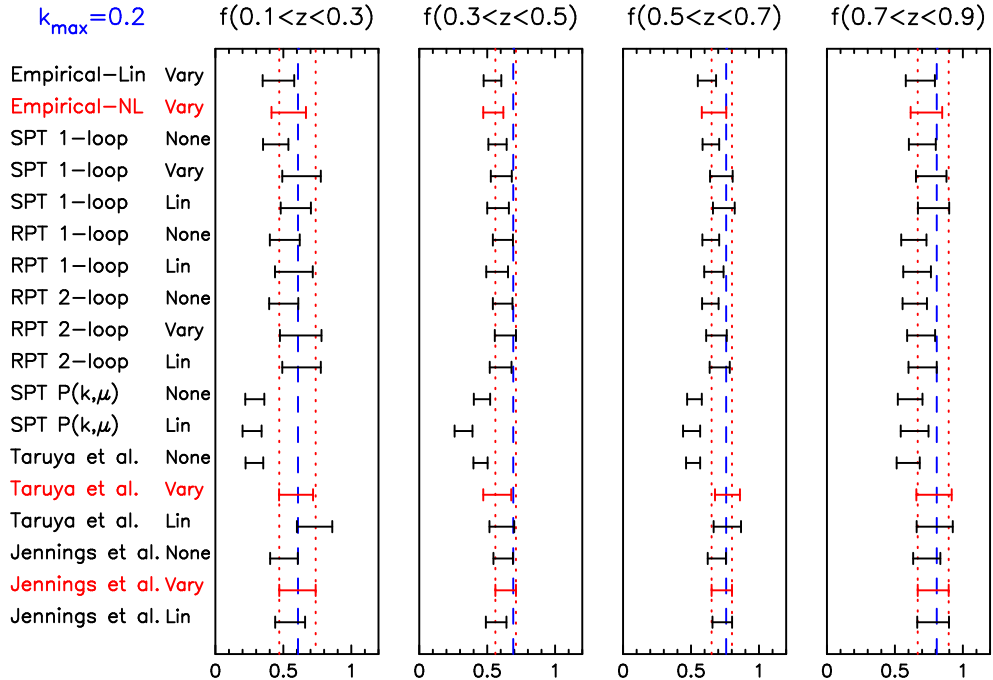}
\caption{Measurements of the growth rate $f$ in four redshift slices
  assuming a fitting limit $k_{\rm max} = 0.2 \, h$ Mpc$^{-1}$ for
  each of the 18 models listed in Table \ref{tabmodlist}.  The
  vertical dashed line indicates the prediction of a flat $\Lambda$CDM
  cosmological model with $\Omega_{\rm m} = 0.27$.  The two vertical
  dotted lines span the $68\%$ confidence region of the growth rate
  measured for the Jennings et al.\ model with a variable damping
  parameter, facilitating an easy comparison of the results for
  different models.  The three best-performing models (based on the
  values of the $\chi^2$ statistic) are highlighted by red text.}
\label{figfmod2}
\end{figure*}

Considering all four redshift slices, the best-performing models for
$k_{\rm max} = 0.3 \, h$ Mpc$^{-1}$ are the Taruya et al.\ (2010)
model, incorporating extra angle-dependent correction terms in
addition to the density and velocity power spectra from 2-loop
Renormalized Perturbation Theory ({\it Model 14} in Table
\ref{tabmodlist}), and the Jennings et al.\ (2011) fitting formula
calibrated from N-body simulations {\it (Model 17)}.  The growth rates
deduced from these two very different modelling techniques agree
remarkably well, after marginalizing over the variable damping term
and linear galaxy bias, with the difference in values being much
smaller than the statistical errors in the measurement.  The level of
this agreement gives us confidence that our results are not limited by
systematic errors.  We note that the empirical Lorentzian streaming
model, where we use the non-linear model power spectrum, also performs
well ({\it Model 2} in Table \ref{tabmodlist}).  In Figure
\ref{figfmod} we have highlighted these three models in red.  For all
scales and redshifts these models typically produce mutually
consistent measurements of the growth rate and minimum values of
$\chi^2$ which differ by $\Delta \chi^2 \sim 1$.  As a further
comparison, Figure \ref{figfmod2} illustrates the measurements for all
18 models in every redshift slice for $k_{\rm max} = 0.2$, again
highlighting the same three optimal models in red.

We can use the dispersion in the results of fitting these three models
to estimate the systematic error in the growth rate measurement, by
taking the variance of the different growth rates weighting by
$\exp{(-\chi^2/2)}$.  The systematic error in $f$ calculated in this
manner is $(0.01, 0.04, 0.03, 0.04)$ in the four redshift slices (for
$k_{\rm max} = 0.3 \, h$ Mpc$^{-1}$).  The magnitude of this error is
less than half that of the statistical error in each bin.  This
systematic error represents the dispersion of growth rate
determinations within the set of redshift-space distortion models
listed in Table \ref{tabmodlist}.

We quote our final results using the Jennings et al.\ (2011)
model, which usually produces the lowest value of $\chi^2$, applied to
$k_{\rm max} = 0.3 \, h$ Mpc$^{-1}$.  The growth rate measurements in
the four redshift slices using this model, marginalizing over the
other parameters, are $f = (0.60 \pm 0.10, 0.70 \pm 0.07, 0.73 \pm
0.07, 0.70 \pm 0.08)$.  The values of the galaxy bias parameter in
each redshift slice using this model, marginalizing over the other
parameters, are $b^2 = (0.69 \pm 0.04, 0.83 \pm 0.04, 1.21 \pm 0.04,
1.48 \pm 0.08)$.

\begin{figure*}
\center
\epsfig{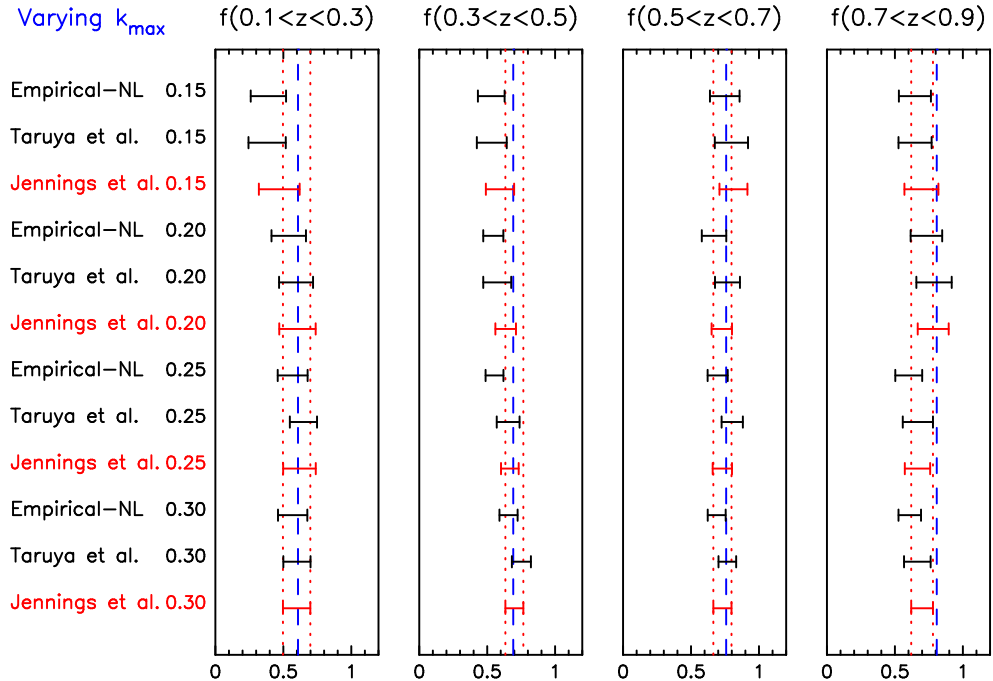}
\caption{Measurements of the growth rate $f$ in four redshift slices,
  varying the fitting limit $k_{\rm max}$ from $0.15$ to $0.3 \, h$
  Mpc$^{-1}$ in steps of $0.05$, for the three optimal models: the
  non-linear empirical Lorentzian, Taruya et al.\ (2010) and Jennings
  et al.\ (2011) models.  All models include a variable damping
  parameter.  The vertical dashed line indicates the prediction of a
  flat $\Lambda$CDM cosmological model with $\Omega_{\rm m} = 0.27$.
  The two vertical dotted lines span the $68\%$ confidence region of
  the growth rate measured for the Jennings et al.\ model for $k_{\rm
    max} = 0.3$, facilitating an easy comparison of the results for
  different models.}
\label{figfmod3}
\end{figure*}

Figure \ref{figfmod3} explores in more detail the robustness of the
growth rate measurements as a function of $k_{\rm max}$ for the three
optimal models.  We plot the growth rate determined in four redshift
slices for these models alone, considering a range of fitting limits
between $k_{\rm max} = 0.15$ and $0.3 \, h$ Mpc$^{-1}$.  Figure
\ref{figfmod3} empirically demonstrates that systematic trends in the
growth rate measurement as $k_{\rm max}$ changes are typically less
than the statistical error in the measurement for $k_{\rm max} = 0.3$.

\begin{figure*}
\center
\epsfig{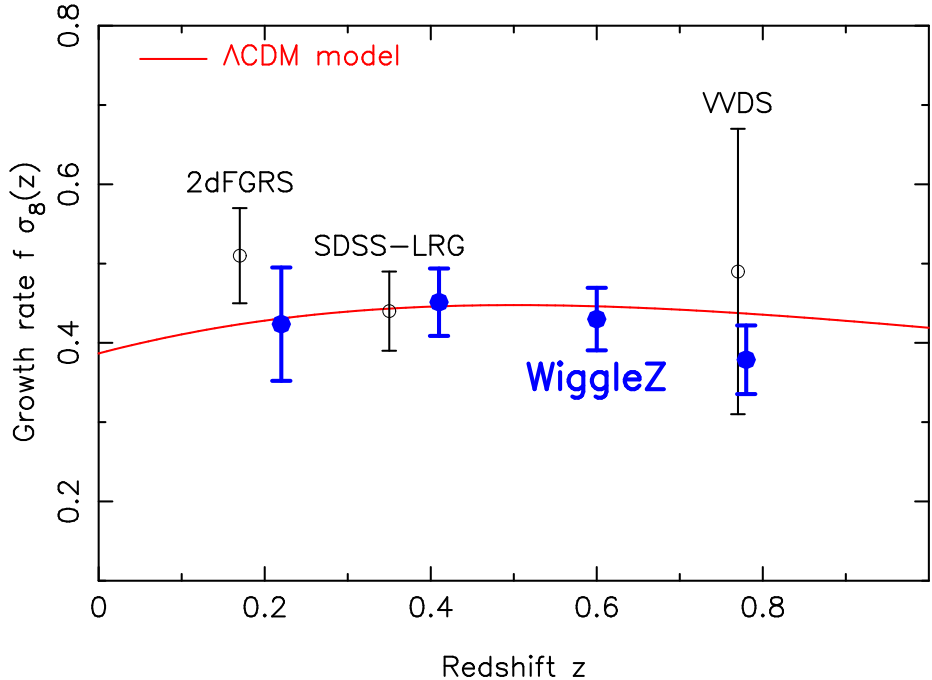}
\caption{Measurements of the growth rate of structure weighted by a
  redshift-dependent normalization, $f(z) \, \sigma_8(z)$, obtained in
  four redshift slices by fitting WiggleZ Survey data.  We assume the
  Jennings et al.\ (2011) model for non-linear redshift-space
  distortions, with a variable damping parameter, and fit to the scale
  range $k < 0.3 \, h$ Mpc$^{-1}$.  The WiggleZ measurements are
  compared to results previously published for the 2dFGRS, SDSS-LRG,
  and VVDS samples (black open circles) as collected by Song \&
  Percival (2009).  The prediction of a flat $\Lambda$CDM cosmological
  model with $\Omega_{\rm m} = 0.27$ is also shown.}
\label{figfsig8}
\end{figure*}

Figure \ref{figfsig8} displays the WiggleZ Survey measurements of the
growth rate of structure in four redshift slices, using the Jennings
et al.\ (2011) model with a variable damping parameter and fitting to
$k_{\rm max} = 0.3 \, h$ Mpc$^{-1}$.  We present our results
multiplied by a redshift-dependent normalization, $f(z) \,
\sigma_8(z)$, where $\sigma_8(z)$ is the r.m.s.\ fluctuation at
redshift $z$ of the linear matter density field in co-moving $8 \,
h^{-1}$ Mpc spheres, calculated for our fiducial cosmological model.
This weighting increases the model-independence of the results by
removing the sensitivity to the overall normalization of the power
spectrum model (Song \& Percival 2009).  Because the overall galaxy
power spectrum amplitude scales with $\sigma_8(z) \, b(z)$ at a
particular redshift $z$, where $b(z)$ is the linear bias factor, and
the magnitude of the redshift-space distortion due to coherent flows
depends on $f(z)/b(z)$, then the measured value of growth rate $f(z)$
scales as $1/\sigma_8(z)$.  The weighted fits in the four redshift
slices are $f(z) \, \sigma_8(z) = (0.42 \pm 0.07, 0.45 \pm 0.04, 0.43
\pm 0.04, 0.38 \pm 0.04)$.  The WiggleZ measurements are compared to
results previously published for the 2dFGRS, SDSS-LRG and VVDS
samples, as collected by Song \& Percival (2009), and to the
prediction of a flat $\Lambda$CDM cosmological model with $\Omega_{\rm
  m} = 0.27$.  We note that:
\begin{itemize}
\item The WiggleZ Survey dataset is the first to produce precise
  growth-rate measurements in the intermediate-redshift range $z >
  0.4$, the apparent transition epoch from decelerating to
  accelerating expansion, with $10\%$ measurement errors that are
  comparable to those obtained at lower redshift from existing
  surveys.
\item The low-redshift $z < 0.4$ WiggleZ measurements agree well with
  existing data.
\item Our dataset permits coherent flows to be quantified across the
  entire redshift range $z < 1$ using observations from a single
  galaxy survey.
\item A cosmological model in which General Relativity describes the
  large-scale gravitation of the Universe, and the current matter
  density parameter is $\Omega_{\rm m} = 0.27$, provides a good
  simultaneous description of the initial conditions described by CMB
  observations, the cosmic expansion history mapped by high-redshift
  supernovae and baryon acoustic oscillations, and the growth history
  mapped by galaxy bulk flows in the WiggleZ Dark Energy Survey.
\end{itemize}

\subsection{Galaxy-mass cross-correlation}

In order to characterize the galaxy bias relation in more detail we
introduced a cross-correlation parameter $r$ between the galaxy and
matter overdensities such that $\langle \delta_g \delta \rangle = b r
\langle \delta^2 \rangle$ and $\langle \delta_g^2 \rangle = b^2
\langle \delta^2 \rangle$ (where $|r| \le 1$ is required by the
definition of a cross-correlation coefficient).  The value $r=1$
corresponds to a fully deterministic bias, whereas $r \le 1$
introduces a random stochastic element to the bias relation.
Measurements of this stochasticity in the SDSS were presented by
Swanson et al.\ (2008), who utilized a counts-in-cells analysis to
quantify its dependence on scale, luminosity and colour.  Swanson et
al.\ found that a scale-independent deterministic linear bias was in
general a good match to the SDSS data, especially on large scales,
where the amplitude of the bias varied significantly with luminosity
for red galaxies but not blue galaxies.  Furthermore, colour-dependent
stochastic effects were evident at smaller scales.  We can extend this
analysis to higher redshifts using the WiggleZ power spectrum.

Equation \ref{eqpklin} may be re-written for a general
cross-correlation parameter $r$ as
\begin{equation}
P_g(k,\mu) = b^2 P_{\delta\delta}(k) - 2 \mu^2 b r P_{\delta\theta}(k)
+ \mu^4 P_{\theta\theta}(k) \; ,
\label{eqpkr}
\end{equation}
and assuming a model for the three power spectra
$P_{\delta\delta}(k)$, $P_{\delta\theta}(k)$ and
$P_{\theta\theta}(k)$, the value of $r$ may be extracted for each
scale $k$ by marginalizing over $b$.  In this investigation we fix the
value of the growth rate $f$ at the value predicted by the
$\Lambda$CDM model, and we assume the Smith et al.\ (2003) and
Jennings et al.\ (2011) prescriptions for the density and velocity
power spectra.  We also marginalized over a variable damping
parameter.

\begin{figure}
\center
\epsfig{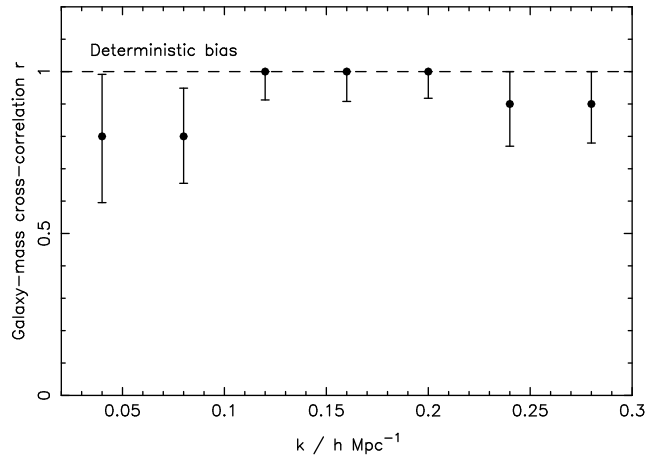}
\caption{The galaxy-mass cross-correlation parameter $r$ as a function
  of scale $k$, measured by fitting Equation \ref{eqpkr} to the
  WiggleZ power spectrum data assuming the growth rate predicted by
  $\Lambda$CDM and marginalizing over linear bias and variable damping
  factors.  The measurements in different redshift slices are
  combined.}
\label{figrvsk}
\end{figure}

Figure \ref{figrvsk} displays the measurement of $r$ in independent
Fourier bins of width $\Delta k = 0.04 \, h$ Mpc$^{-1}$ between $k =
0.02$ and $0.3 h$ Mpc$^{-1}$, combining the results for different
redshift slices and varying $r$ within the range $-1 \le r \le 1$.  We
find that the cross-correlation parameter is consistent with
deterministic bias $r=1$ (and this result also applies for each
separate redshift slice).  Because the probability distribution for
$r$ is asymmetric due to the hard upper limit, in the cases when the
confidence region is truncated at $r=1$ we plot in Figure
\ref{figrvsk} the range below $r=1$ enclosing $68\%$ of the
probability, and the position of the peak of the likelihood.

\section{Analysis of the power spectrum moments}
\label{secpkmom}

\subsection{Multipole moments of the power spectrum}

In this Section we explore some alternative techniques for quantifying
the redshift-space power spectra which can visualize their
information content more neatly.  The galaxy power spectrum
$P_g^s(k,\mu)$ may be decomposed in a basis of Legendre polynomials
$L_\ell(\mu)$ to give multipole moments $P_\ell(k)$:
\begin{eqnarray}
P_g^s(k,\mu) &=& \sum_{{\rm even} \, \ell} P_\ell(k) \, L_\ell(\mu) \\
P_\ell(k) &=& \frac{2\ell + 1}{2} \int_{-1}^{1} d\mu \, P_g^s(k,\mu) \, L_\ell(\mu) \; .
\label{eqpkl}
\end{eqnarray}
The monopole ($l=0$) spectrum primarily contains information about the
underlying shape of the isotropic clustering pattern.  The quadrupole
($l=2$) spectrum holds the leading-order signal from the anisotropic
modulation in power due to redshift-space effects.  We note that the
multipole moments may be expressed in terms of the density-velocity
power spectra $P_{\delta\delta}$, $P_{\delta\theta}$ and
$P_{\theta\theta}$ (e.g.\ Percival \& White 2009).

The multipole moments may be extracted from the power spectrum
measurement in bins of $\mu$ by turning Equation \ref{eqpkl} from an
integral into a sum:
\begin{equation}
P_\ell(k) = \frac{2\ell + 1}{2} \sum_{\mu \, {\rm bins}} P_g^s(k,\mu) \int_{\mu-\Delta \mu/2}^{\mu+\Delta \mu/2} L_\ell(\mu') \, d\mu' \; .
\label{eqpklest1}
\end{equation}
Alternatively, Yamamoto et al.\ (2006) introduced a direct estimator
for $P_\ell(k)$ which does not require binning of the power spectrum
in $\mu$ (which is particularly problematic at low $k$, where there
are limited modes available in Fourier space).  We present the key
equations of the estimator here, referring the reader to Yamamoto et
al.\ (2006) for the full derivation.

The Yamamoto et al.\ estimator, which is valid when the
distant-observer approximation is applicable, is written using sums
over $N_{\rm gal}$ observed galaxies and $N_{\rm ran} = N_{\rm
  gal}/\alpha$ random (mock) galaxies (where $\alpha \ll 1$).  For
each Fourier mode $\vec{k}$ we define the multipole moments based on
the data as
\begin{equation}
D_\ell(\vec{k}) = \sum_{i=1}^{N_{\rm gal}} w(\vec{s}_i) \, \exp{(i
  \vec{s}_i.\vec{k})} \, L_\ell(\hat{\vec{s}}_i . \hat{\vec{k}}) \; ,
\end{equation}
where $\vec{s}_i$ is the position vector of galaxy $i$ and
$w(\vec{s})$ is a weight factor for each galaxy, specified below.  If
we define the equivalent sum $R_\ell(\vec{k})$ over the set of random
galaxies, then an estimator for $P_\ell(\vec{k})$ is
\begin{eqnarray}
P_\ell(\vec{k}) &=& A^{-1} \left[ D_\ell(\vec{k}) - \alpha R_\ell(\vec{k}) \right] \left[ D_0(\vec{k}) - \alpha R_0(\vec{k}) \right] \nonumber \\
&-& S_\ell(\vec{k}) \; ,
\label{eqpklest2}
\end{eqnarray}
where the shot noise term $S_\ell(\vec{k})$ is given by
\begin{equation}
S_\ell(\vec{k}) = A^{-1} (1+\alpha) \alpha \sum_{i=1}^{N_{\rm ran}} w(\vec{s}_i)^2 \, L_\ell(\hat{\vec{s}}_i . \hat{\vec{k}}) \; .
\end{equation}
The normalization $A$ is given, in terms of sums over the $N_c$ grid
cells $\vec{x}$ constituting the window function, as
\begin{equation}
A = \sum_{\vec{x}} W^2(\vec{x}) w^2(\vec{x}) = \sum_{i=1}^{N_{\rm gal}} W(\vec{s}_i) \, w^2(\vec{s}_i) \; ,
\end{equation}
where $W(\vec{s})$ is the window function normalized over its volume
$V$ such that $\int W dV = N_{\rm gal}$, or $\sum_{\vec{x}} W(\vec{x})
= N_{\rm gal} (N_c/V)$.  The minimum variance in $P_\ell(\vec{k})$ is
produced by the usual FKP weight function
\begin{equation}
w(\vec{s}) = [ 1 + W(\vec{s}) P_0 ]^{-1} \; ,
\end{equation}
where $P_0$ is a characteristic power spectrum amplitude (which we
take as $P_0 = 5000 \, h^{-3}$ Mpc$^3$, although this choice has very
little effect on our results).  The error in the estimator for each
Fourier mode $\vec{k}$ is given by
\begin{eqnarray}
\left[ \Delta P_\ell(\vec{k}) \right]^2 &=& A^{-1} \alpha
\sum_{i=1}^{N_{\rm ran}} w(\vec{s}_i)^4 W(\vec{s}_i) \nonumber \\
&\times& \left[ W(\vec{s}_i) P(\vec{k}) + 1+\alpha \right]^2 \left[
  L_\ell(\hat{\vec{s}}_i . \hat{\vec{k}}) \right]^2 \; .
\end{eqnarray}
We evaluated the estimator for $P_\ell(\vec{k})$ over the usual grid
of Fourier modes which describe fluctuations in a cuboid of dimensions
$(L_x,L_y,L_z)$, i.e.\ for modes $\vec{k} = (k_x,k_y,k_z) = (2\pi
n_x/L_x, 2\pi n_y/L_y, 2\pi n_z/L_z)$ for integers $(n_x,n_y,n_z)$.
We then averaged the amplitudes in spherical shells of $\vec{k}$ to
produce our estimate of $P_\ell(k)$, which we write as $P_\ell^{\rm
  gridded}(k)$.  We note that the discreteness of the Fourier modes in
the grid produces a bias in the estimate, which is particularly
evident at low $k$.  We corrected for this bias using a model power
spectrum $P^{\rm model}(\vec{k})$ by evaluating
\begin{eqnarray}
P_\ell^{\rm model,gridded}(\vec{k}) &=& A^{-1} \alpha
\sum_{i=1}^{N_{\rm ran}} w(\vec{s}_i)^2 W(\vec{s}_i) \nonumber \\
&\times& P^{\rm model}(\vec{k}) L_\ell(\hat{\vec{s}}_i
. \hat{\vec{k}}) \; ,
\end{eqnarray}
which we averaged in spherical shells of $k$ to produce $P_\ell^{\rm
  model,gridded}(k)$, and also an exact determination using
\begin{equation}
P_\ell^{\rm model,exact}(k) = \frac{2\ell+1}{2} \int_{-1}^1 d\mu \,
P^{\rm model}(k,\mu) \, L_\ell(\mu) \; .
\end{equation}
Our final estimate for the multipole power spectrum is then given by
\begin{equation}
P_\ell(k) = P_\ell^{\rm gridded}(k) + P_\ell^{\rm model,exact} -
P_\ell^{\rm model,gridded} \; .
\end{equation}
We generated this correction using the best-fitting non-linear
empirical Lorentzian redshift-space power spectrum (see Section
\ref{secempmodel}) as the input model $P^{\rm model}(\vec{k})$.

\begin{figure*}
\center
\epsfig{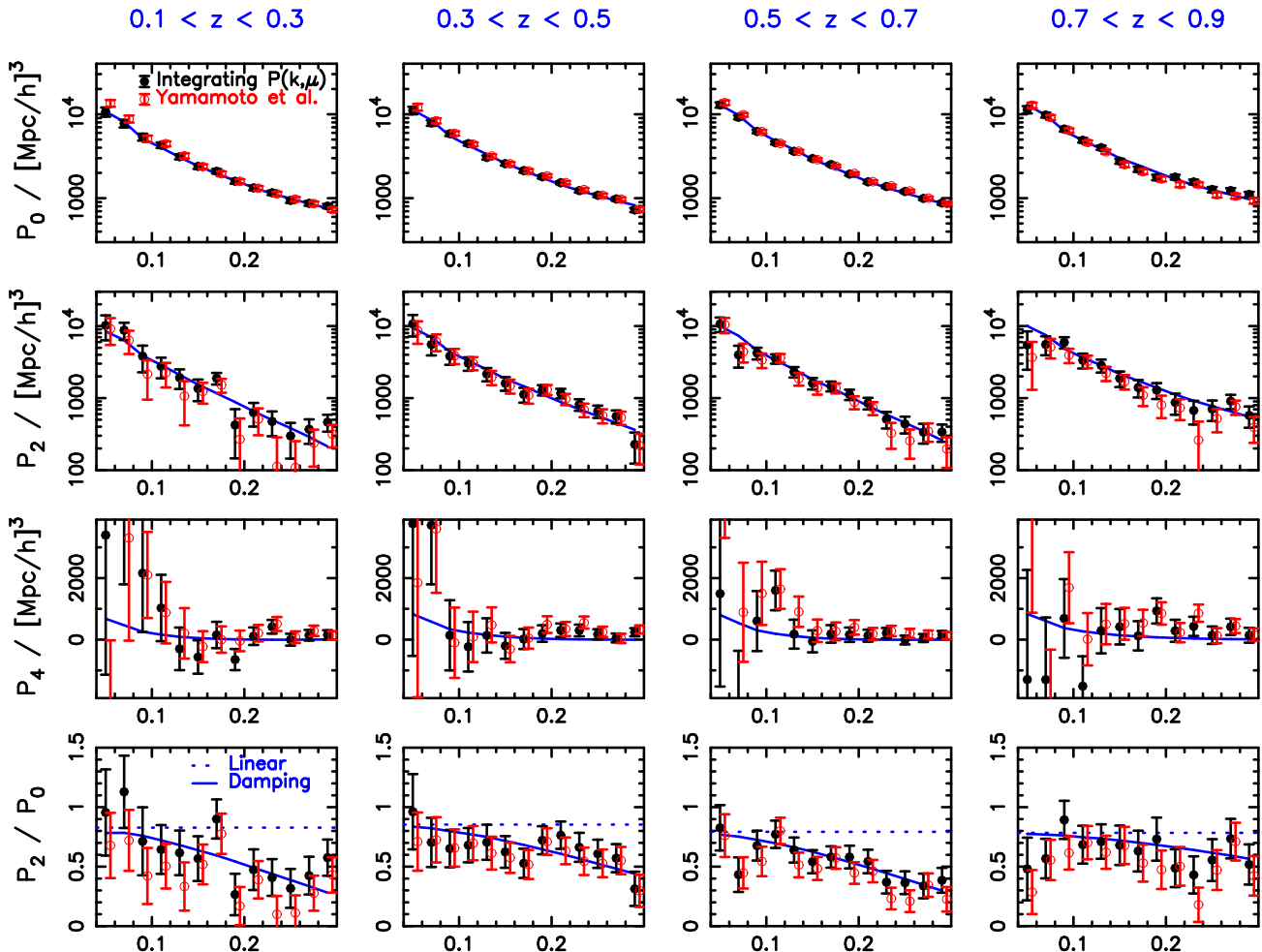}
\caption{The multipole power spectra $P_\ell(k)$ for $\ell=0,2,4$ for
  WiggleZ survey observations in four redshift slices.  The monopole
  ($l=0$) spectrum primarily contains information about the underlying
  shape of the isotropic clustering pattern.  The quadrupole ($l=2$)
  spectrum holds the leading-order signal from the anisotropic
  modulation in power due to redshift-space effects.  The (black)
  solid circles are generated from the stacked measurements of
  $P_g(k,\mu)$ across the different survey regions using Equation
  \ref{eqpklest1}.  The (red) open circles, which are offset slightly
  in the $x$-direction for clarity, are generated by combining the
  estimates of $P_\ell(k)$ in each region using the Yamamoto et
  al.\ estimator of Equation \ref{eqpklest2}.  The model lines
  correspond to the best-fitting non-linear empirical Lorentzian
  redshift-space distortion model in each case.  The bottom row
  displays the quadrupole-to-monopole ratio $P_2(k)/P_0(k)$.  Two
  models are overplotted: the large-scale Kaiser limit predicted in a
  $\Lambda$CDM cosmological model with $\Omega_{\rm m} = 0.27$,
  labelled as ``Linear'', and the non-linear redshift-space distortion
  model, labelled as ``Damping''.}
\label{figpklmeas}
\end{figure*}

Figure \ref{figpklmeas} compares the measurement of the multipole
power spectra in four redshift slices obtained by the direct sum of
Equation \ref{eqpklest1} with the Yamamoto et al.\ estimator of
Equation \ref{eqpklest2}.  In general the two different techniques for
deriving the multipole power spectra agree well and we obtain
measurements of the monopole ($\ell = 0$) and quadrupole ($\ell = 2$)
with high signal-to-noise.  Current galaxy redshift surveys do not
yield a significant detection of the hexadecapole ($\ell = 4$).

The final row of Figure \ref{figpklmeas} plots the measured
quadrupole-to-monopole ratio $P_2(k)/P_0(k)$ as a function of scale
for each redshift slice.  This statistic has the advantage of being
less sensitive than the power spectrum itself to the parameters which
model the shape of the underlying real-space galaxy clustering pattern
(such as the background cosmological parameters or a scale-dependent
bias).  On large scales this ratio is expected to asymptote to a
constant value which may be derived from Equation \ref{eqpkkaiser}:
\begin{equation}
\frac{P_2(k)}{P_0(k)} = \frac{\frac{4}{3} \beta + \frac{4}{7}
  \beta^2}{1 + \frac{2}{3} \beta + \frac{1}{5} \beta^2} \; ,
\end{equation}
where $\beta = f/b$.  This value, indicated by the dotted ``Linear''
horizontal line in the bottom row of Figure \ref{figpklmeas}, and
derived using the prediction of $f(z)$ in a $\Lambda$CDM model with
$\Omega_{\rm m} = 0.27$, lies in good agreement with the measurements
on large scales in each redshift slice.  At smaller scales the data
deviates from this prediction due to the non-linear effects which damp
the velocity power spectrum.  We also plot the scale-dependent value
of $P_2(k)/P_0(k)$ for the best-fitting non-linear empirical
Lorentzian redshift-space distortion model in each redshift slice,
indicated by the ``Damping'' line.

\subsection{Power spectra of the velocity divergence field}

The characteristic angular dependence of the redshift-space galaxy
power spectrum $P_g^s(k,\mu)$ on its three component power spectra
$P_{gg}(k)$, $P_{g\theta}(k)$ and $P_{\theta\theta}(k)$, exhibited by
Equation \ref{eqpklin}, gives us the opportunity to extract these
three power spectra directly from data.  This is of particular
interest for the case of $P_{\theta\theta}(k)$ because this quantity
depends on the growth rate but not on the galaxy bias, which is
considered to be one of the principle sources of potential systematic
error in redshift-space distortion model-fitting.

The signal-to-noise ratio of the power spectrum measurements from
current surveys is not yet sufficiently high to extract three
independent functions cleanly (e.g., Tegmark et al.\ 2004) -- which is
consistent with our failure to detect the hexadecapole in Figure
\ref{figpklmeas}.  However, a good approximation of the
galaxy-velocity cross-power spectrum in the quasi-linear regime is
$P_{g\theta} = -\sqrt{P_{gg} P_{\theta\theta}}$ (Percival \& White
2009), which cancels (to first order) non-linear terms in the power
spectra and galaxy bias.  Under this approximation we can fit for the
coefficients $P_{gg}(k)$ and $P_{\theta\theta}(k)$ in the model (Song
\& Kayo 2010)
\begin{equation}
P_g^s(k,\mu) = P_{gg}(k) + 2 \mu^2 \sqrt{P_{gg}(k)
  P_{\theta\theta}(k)} + \mu^4 P_{\theta\theta}(k) \; .
\label{eqpkggvv}
\end{equation}
For each separate $k$-bin, spaced by $\Delta k = 0.02 \, h$
Mpc$^{-1}$, we fitted the model of Equation \ref{eqpkggvv} to the
stacked measurements of $P_g(k,\mu)$ from the WiggleZ survey dataset
in four redshift slices.  We performed the fit in 10 Fourier bins up
to $k_{\rm max} = 0.2 \, h$ Mpc$^{-1}$, choosing this upper limit
because Equation \ref{eqpklin} will likely not provide a reliable
description of the $\mu$-dependence of the power spectrum at smaller
scales (given that our model fits in Section \ref{secgrowthfit} favour
the inclusion of an additional Lorentzian damping term over the range
$0.2 < k < 0.3 \, h$ Mpc$^{-1}$).

\begin{figure*}
\center
\epsfig{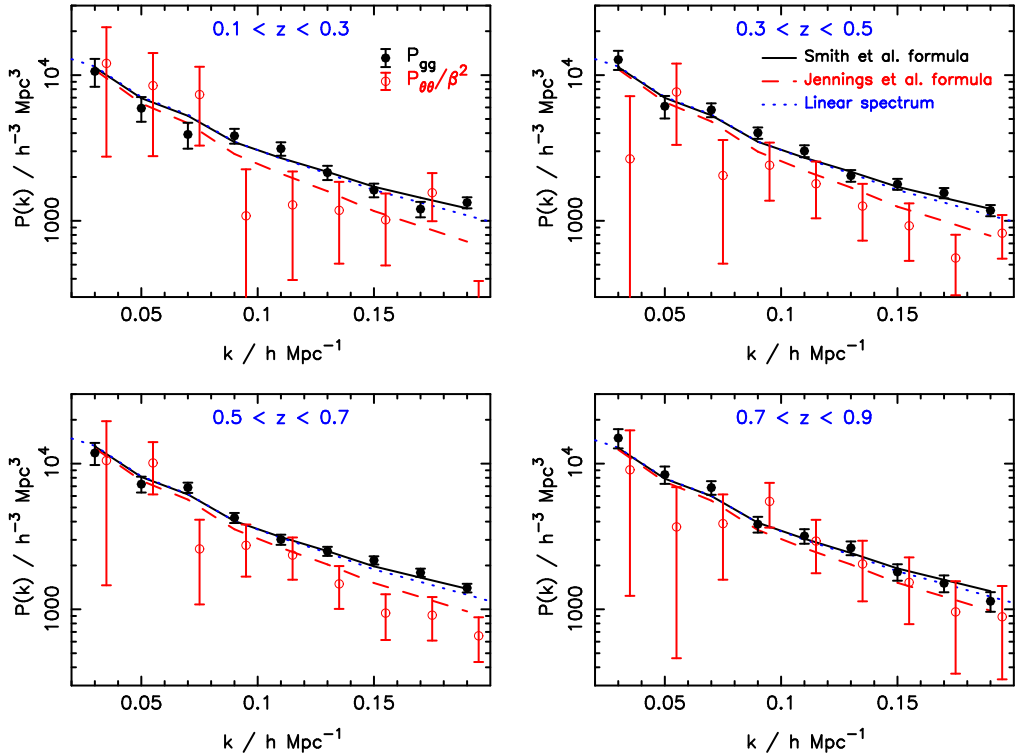}
\caption{Measurement of the WiggleZ survey galaxy-galaxy and
  velocity-velocity power spectra in four redshift slices by
  maximum-likelihood fitting to the stacked measurements of
  $P_g(k,\mu)$ across the different survey regions using the model of
  Equation \ref{eqpkggvv}.  The measurements of $P_{\theta\theta}$ are
  normalized by $\beta^{-2}$ (in order to match the large-scale limit
  of $P_{gg}$) and are offset slightly in the $x$-direction for
  clarity.  For comparison, we plot the linear-regime matter power
  spectra, the non-linear matter power spectra from Smith et
  al.\ (2003) and the non-linear velocity power spectra from Jennings
  et al.\ (2011).  Our extraction of these two power spectra rests on
  the assumption that $P_{g\theta} = -\sqrt{P_{gg} P_{\theta\theta}}$
  (Song \& Kayo 2010), which has been validated at large scales by
  simulations (Percival \& White 2009).}
\label{figpkggvv}
\end{figure*}

Figure \ref{figpkggvv} displays the results of the fits for each
redshift slice, where for convenience we have divided the measurements
of $P_{\theta\theta}(k)$ by the best-fitting value of $\beta^2 =
(f/b)^2$ so that the galaxy and velocity power spectra are expected to
have the same large-scale limit.  For comparison we also plot in each
case the non-linear galaxy and velocity power spectra based on the
fitting formulae proposed by Smith et al.\ (2003) and Jennings et
al.\ (2011), respectively, together with the underlying linear matter
power spectrum for our fiducial cosmological parameters at these
redshifts.

Our measurements constitute the first determination of the velocity
power spectrum as a function of redshift, and cleanly reveal the
effects that we are modelling.  At large scales $k < 0.1 \, h$
Mpc$^{-1}$ the density and velocity power spectra are in close
agreement with each other and the input model linear power spectrum.
At smaller scales the measurements diverge: the density power spectra
are boosted in amplitude in a manner that closely matches the fitting
formula of Smith et al.\ (2003), and the velocity power spectra are
damped by non-linear effects.  The fitting formula of Jennings et
al.\ (2011) provides a good match to this damping: the value of
$\chi^2$ statistic is $(13.4, 11.8, 12.4, 3.6)$ for the four redshift
slices respectively, for 10 degrees of freedom.  The value of $\chi^2$
for the highest redshift bin corresponds to a 2-$\sigma$ fluctuation.
As the $\chi^2$ values for the other three redshift slices fall within
the 1-$\sigma$ range for the distribution, we do not view this with
concern.  We also find a tentative indication that the amplitude of
the non-linear correction to the velocity power spectrum increases
with decreasing redshift, expected as a consequence of the growth of
structure.

\section{Conclusions}
\label{secconc}

We have used the WiggleZ Dark Energy Survey dataset to produce the
first precise map of cosmic growth spanning the epoch of cosmic
acceleration and the first systematic study of the growth history from
a single galaxy survey.  We have compared the measured power spectra
to 18 different redshift-space distortion models using a combination
of empirical models, fitting formulae calibrated by N-body
simulations, and perturbation theory techniques.  We itemize our
conclusions as follows:

\begin{itemize}

\item Two quasi-linear redshift-space distortion models provide a good
  description of our data for scales $k < 0.3 \, h$ Mpc$^{-1}$: the
  Taruya et al.\ (2010) model, incorporating extra angle-dependent
  correction terms in addition to the density and velocity power
  spectra from 2-loop Renormalized Perturbation Theory, and the
  Jennings et al.\ (2011) fitting formula calibrated from N-body
  simulations.  In each model we included a variable damping
  parameter.  The growth rates deduced from these two very different
  modelling techniques agree remarkably well, with the difference in
  values being much smaller than the statistical errors in the
  measurement.  The level of this agreement gives us confidence that
  our results are not limited by systematic errors.  We note that the
  empirical Lorentzian streaming model, where we use the non-linear
  matter power spectrum from Smith et al.\ (2003), also performs well
  and the minimum chi-squared values for these three models typically
  differ by $\Delta \chi^2 \approx 1$.  We quote our final results
  using the Jennings et al.\ (2011) model, which usually produces the
  lowest value of $\chi^2$: growth rate measurements of $f(z) = (0.60
  \pm 0.10, 0.70 \pm 0.07, 0.73 \pm 0.07, 0.70 \pm 0.08)$ at redshifts
  $z = (0.22, 0.41, 0.6, 0.78)$, where we have marginalized over the
  variable damping factor and a linear galaxy bias factor.  A more
  model-independent way of expressing these fits is $f(z) \,
  \sigma_8(z) = (0.42 \pm 0.07, 0.45 \pm 0.04, 0.43 \pm 0.04, 0.38 \pm
  0.04)$.

\item These growth rate measurements are consistent with those
  expected in a flat General Relativistic $\Lambda$CDM cosmological
  model with matter density $\Omega_{\rm m} = 0.27$.  Our observations
  therefore indicate that this model provides a self-consistent
  description of the growth of cosmic structure from perturbations and
  the large-scale, homogeneous cosmic expansion mapped by supernovae
  and baryon acoustic oscillations.

\item Assuming the growth rate predicted by the $\Lambda$CDM model we
  can fit for the parameters of a stochastic scale-dependent bias
  described by a galaxy-mass cross-correlation $r(k)$.  We find that
  this bias is consistent with a deterministic model $r=1$ for the
  range of scales $k < 0.3 \, h$ Mpc$^{-1}$.

\item We considered various methods for presenting the information
  contained in the redshift-space power spectra, including deriving
  the multipole moments $P_\ell(k)$ using direct integration of the
  binned power spectrum $P(k,\mu)$ and by implementing the estimator
  described by Yamamoto et al. (2006).  Measurements of the
  quadrupole-to-monopole ratio $P_2/P_0$ as a function of scale $k$
  delineate the influence of redshift space distortions in a manner
  independent of the shape of the underlying matter power spectrum or
  a scale-dependent bias.

\item Under the assumption $P_{g\theta} = -\sqrt{P_{gg}
  P_{\theta\theta}}$, which is a good approximation in the
  quasi-linear regime, we used the redshift-space power spectra to fit
  directly for $P_{gg}(k)$ and $P_{\theta\theta}(k)$.  We found that
  (within an overall normalization factor) the galaxy and velocity
  power spectra are consistent with each other and with the model
  linear power spectrum at low $k$.  For $k > 0.1 \, h$ Mpc$^{-1}$ we
  delineated for the first time the characteristic non-linear damping
  of the velocity power spectrum as a function of redshift, with a
  tentative indication that the amplitude of the non-linear effects
  increases with decreasing redshifts.  The Jennings et al.\ (2011)
  fitting formula provides a good fit to these power spectra.

\end{itemize}

A future investigation will involve the confrontation of this data
with a range of modified-gravity models, combining the large-scale
structure measurements with self-consistent fits to the Cosmic
Microwave Background observations.  Furthermore, a joint analysis of
the redshift-space distortions and Alcock-Paczynski effect is also in
preparation.

\section*{Acknowledgments}

We thank Carlton Baugh, Elise Jennings, Juliana Kwan, David Parkinson,
Will Percival, Roman Scoccimarro and Yong-Seon Song for useful
comments which influenced and improved the development of this paper.
We are particularly grateful to Martin Crocce for providing power
spectra for Renormalized Perturbation Theory and for helpful comments.

We acknowledge financial support from the Australian Research Council
through Discovery Project grants funding the positions of SB, MP, GP
and TD.  SMC acknowledges the support of the Australian Research
Council through a QEII Fellowship.  MJD and TD thank the Gregg
Thompson Dark Energy Travel Fund for financial support.

GALEX (the Galaxy Evolution Explorer) is a NASA Small Explorer,
launched in April 2003.  We gratefully acknowledge NASA's support for
construction, operation and science analysis for the GALEX mission,
developed in co-operation with the Centre National d'Etudes Spatiales
of France and the Korean Ministry of Science and Technology.

Finally, the WiggleZ survey would not be possible without the
dedicated work of the staff of the Australian Astronomical Observatory
in the development and support of the AAOmega spectrograph, and the
running of the AAT.

\end{document}